%% file: ssing3.tex
\documentclass[12pt, twoside]{article}
\usepackage{amsmath, amsfonts, amscd, amssymb}
\usepackage{amsthm, fancybox}
\usepackage{eufrak, euscript, oldgerm, graphics}

\newcommand{\aconn}{\mathcal{A}}

\newcommand{\aut}{{\mathcal{A}}ut}

\newcommand{\diad}{\mathbf{D}}

\newcommand{\conn}{\mathcal{D}}

\newcommand{\cons}{\mathbf{C}}

\newcommand{\curv}{R}

\newcommand{\ric}{{\mathcal{R}}}
\newcommand{\simp}{\mathcal{K}}

\newcommand{\ricci}{\EuScript{R}}

\newcommand{\eh}{\mathfrak{E}\mathfrak{H}}

\newcommand{\gauge}{\mathcal{U}}

\newcommand{\Hom}{{\mathcal{H}}om}

\newcommand{\morph}{\EuScript{F}}
\newcommand{\inv}{\overleftarrow{\EuScript{P}}}
\newcommand{\inveinst}{\overleftarrow{\EuScript{E}}}
\newcommand{\invmod}{\overleftarrow{\EuScript{M}}}
\newcommand{\invs}{\overleftarrow{\EuScript{S}}}
\newcommand{\invsconn}{\overleftarrow{\mathfrak{A}}}
\newcommand{\invel}{\overleftarrow{\mathfrak{E}}}
\newcommand{\inveh}{\overleftarrow{\EuScript{EH}}}
\newcommand{\invcq}{\overleftarrow{\EuScript{Z}}}
\newcommand{\invtriad}{\stackrel{\rightleftarrows}{\EuScript{T}}}
\newcommand{\finv}{\overleftarrow{{\EuScript{G}}}}
\newcommand{\diromg}{\overrightarrow{\mathfrak{R}}}

\newcommand{\man}{{\mathcal{M}}an}
\newcommand{\modl}{\mathbf{\mathcal{E}}}

\newcommand{\omg}{\Omega}
\newcommand{\Omg}{\mathbf{\Omega}}

\newcommand{\cont}{\mathcal{C}^{0}}
\newcommand{\smooth}{\mathcal{C}^{\infty}}
\newcommand{\anal}{\mathcal{C}^{\omega}}
\newcommand{\sconn}{\textsf{A}}%space of general connections
\newcommand{\striad}{\large\texttt{T}_{\infty}}
\newcommand{\ssmooth}{\EuScript{C}^{\infty}}

\newcommand{\struc}{\mathbf{A}}

\newcommand{\triad}{{\mathfrak{T}}}
\newcommand{\ctriad}{{\mathfrak{D}}\triad}

\newcommand{\wee}{\,{\scriptstyle\wedge}\,}
\newcommand{\bull}{{\scriptstyle\bullet}}

\newcommand{\com}{\mathbb{C}}
\newcommand{\mapto}{\longrightarrow}

\newcommand{\Z}{\mathbb{Z}}
\newcommand{\R}{\mathbb{R}}

\input{diagram}

\title{\bf Finitary-Algebraic `Resolution' of\\ the Inner Schwarzschild Singularity\thanks{To appear in its present form at the {\it International Journal of
Theoretical Physics}, {\bf 44}, (12) (2005); gr-qc/0408045.}}

\author{Ioannis Raptis\thanks{EU Marie Curie Reintegration Research Fellow, Algebra and Geometry Section,
Department of Mathematics, University of Athens,
Panepistimioupolis, Athens 157 84, Greece; {\em and} Visiting
Researcher, Theoretical Physics Group, Blackett Laboratory,
Imperial College of Science, Technology and Medicine, Prince
Consort Road, South Kensington, London SW7 2BZ, UK; e-mail:
i.raptis@ic.ac.uk}}

\date{}

\begin{document}

{\catcode`\ =13\global\let =\ \catcode`\^^M=13
\gdef^^M{\par\noindent}}
\def\verbatim{\tt
\catcode`\^^M=13 \catcode`\ =13 \catcode`\\=12 \catcode`\{=12
\catcode`\}=12 \catcode`\_=12 \catcode`\^=12 \catcode`\&=12
\catcode`\~=12 \catcode`\#=12 \catcode`\%=12 \catcode`\$=12
\catcode`|=0 }

\maketitle

\pagestyle{myheadings}\markboth{\centerline {\small {\sc{Ioannis
Raptis}}}}{\centerline {\footnotesize {\sc Finitary-Algebraic
`Resolution' of the Inner Schwarzschild Singularity}}}

\pagenumbering{arabic}

\begin{abstract}

\noindent A `resolution' of the interior singularity of the
spherically symmetric Schwarzschild solution of the Einstein
equations for the gravitational field of a point-particle is
carried out entirely and solely by finitistic and algebraic means.
To this end, the background differential spacetime manifold and,
{\it in extenso}, Differential Calculus-free purely algebraic
(:sheaf-theoretic) conceptual and technical machinery of {\em
Abstract Differential Geometry} (ADG) is employed. As in previous
work \cite{malrap1,malrap2,malrap3}, which this paper continues,
the starting point for the present application of ADG is Sorkin's
finitary (:locally finite) poset (:partially ordered set)
substitutes of continuous manifolds in their Gel'fand-dual picture
in terms of discrete differential incidence algebras and the
finitary spacetime sheaves thereof. It is shown that the Einstein
equations hold not only at the finitary poset level of `discrete
events', but also at a suitable `classical spacetime continuum
limit' of the said finitary sheaves and the associated
differential triads that they define ADG-theoretically. The upshot
of this is twofold: on the one hand, the field equations are seen
to hold when only finitely many events or `degrees of freedom' of
the gravitational field are involved, so that no infinity or
uncontrollable divergence of the latter arises at all in our
inherently finitistic-algebraic scenario. On the other hand, the
law of gravity---still modelled in ADG by a {\em differential}
equation proper---does {\em not} break down in any (differential
geometric) sense in the vicinity of the locus of the point-mass as
it is traditionally maintained in the usual manifold based
analysis of spacetime singularities in General Relativity (GR). At
the end, some brief remarks are made on the potential import of
ADG-theoretic ideas in developing a genuinely background
independent Quantum Gravity (QG). A brief comparison between the
`resolution' proposed here and a recent resolution of the inner
Schwarzschild singularity by Loop QG means concludes the paper.

\vskip 0.2in



\end{abstract}

\newpage  \setlength{\textwidth}{17.0cm} % OK for both Letter and A4
%\setlength{\oddsidemargin}{0cm}  %margins = 1inch +
                                  %top/odd/even-sidemargin
%\setlength{\evensidemargin}{0cm} %  ditto

\oddsidemargin=-.8cm \evensidemargin=-1cm \headsep=.8cm

\setlength{\topskip}{0pt}  %see pp155 also about baselineskip
\setlength{\textheight}{23.2cm} % Use 23cm for Letter
\setlength{\footskip}{-2cm}

\setlength{\topmargin}{0pt}

\newpage

\section{Prolegomena: Introduction cum Physical Motivation}

It is widely maintained today that, given certain broad
assumptions and generic conditions, General Relativity (GR)
`predicts' singularities---{\it loci} in the spacetime continuum
where the gravitational field grows uncontrollably without bound
and, ultimately, the Einstein equations that it obeys `break down'
in one way or another. At the same time however, few physicists
would disagree that the main culprit for these pathologies and
their associated unphysical infinities is {\em our} model of
spacetime as a pointed, $\smooth$-smooth (:differential)
manifold.\footnote{In the present paper we tacitly identify the
physicists' intuitive term `{\em spacetime continuum}' with the
mathematicians' notion of a (finite-dimensional) `{\em locally
Euclidean space}'---{\it ie}, a manifold, looking locally like
$\R^{n}$ and carrying the usual topological ($\cont$) and
differential ($\smooth$) structure.}

Granted that the said anomalies and divergences are physically
unacceptable, but at the same time that the whole conceptual
edifice and technical machinery of Classical Differential Geometry
(CDG)---the mathematical language in which GR is traditionally
formulated\footnote{In the original formulation of GR by Einstein,
CDG pertains to the pseudo-Riemannian geometry of smooth
manifolds.}---vitally depends on a base smooth manifold, the
physicist appears to be impaled on the horns of a dilemma. On the
one hand, she wishes to do away with singularities and their
physically meaningless infinities, while on the other, she wishes
to retain (or anyway, she is reluctant to readily abandon) the
picture of a physical law (here, the law of gravity) as a {\em
differential equation} proper, if anything in order for the theory
still to be able to accommodate some notion of {\em locality}---be
it {\em differential} locality.\footnote{That is, the idea that
dynamical gravitational field actions connect infinitesimally
separated events, or equivalently, that events causally influence
others in their `infinitesimal neighborhood' (differential
locality or local causality in the point-set manifold of events).}

In other words, the tension may be expressed as follows: how can
one get rid of the spacetime continuum with its `inherent'
singularities, but still be able to apply somehow {\em
differential} geometric ideas to theoretical physics? Especially
in GR, this friction manifests itself in the glaring conflict
between the Principle of General Covariance (PGC) and the
fruitless attempts so far at defining {\em precisely} what is a
singularity in the theory \cite{geroch,hawk0,clarke4,rendall}. For
if one does away with the differential manifold model for
spacetime, and, as a result, the whole of the CDG based on it, one
has also got to abandon the by now standard mathematical
representation of the PGC by $\mathrm{Diff}(M)$---the
diffeomorphism `symmetry' group of automorphisms of the underlying
smooth continuum $M$. No matter how easily the theoretical
physicist may be convinced to abandon the {\em mathematical} (and
quite {\it a priori}!) assumption of the spacetime continuum if
the nonsensical singularities have to go with it, she will not be
as easily prepared to sacrifice {\em the} pillar on which GR, as a
{\em physical} theory, stands---the PGC. Otherwise, at least she
is forced to look for an alternative mathematical expression for
it---one that, unlike the traditional one involving
$\mathrm{Diff}(M)$, is {\em not} dictated by the smooth background
manifold.

In order to appreciate how formidable this dilemma-{\it
cum}-impasse is, one has to consider that, arguably, the only way
we actually know how to do {\em differential} geometry is via a
base manifold ({\it ie}, CDG); albeit, in doing CDG we have to put
up with the singularities that are built into $M$. Let it be
stressed here that we tacitly assume that {\em a differential
manifold} $M$ {\em is nothing else but the algebra} $\smooth(M)$
{\em of infinitely differentiable `coordinate' functions labelling
its points} (Gel'fand duality/spectral theory)
\cite{malrap2,malrap3,malrap4}. Thus, when we say that
singularities are `inherent' or built into $M$, we mean that they
are singularities of some smooth function in
$\smooth(M)$.\footnote{For example, the algebra of coordinates in
which $g_{\mu\nu}$---{\em the} principal dynamical variable in GR,
whose components represent the ten gravitational
potentials---takes its values, is $\smooth(M)$. That is, the said
decade of potentials are {\em smooth} functions on $M$, and
precisely because of this one says that the metric tensor
$g_{\mu\nu}$ is a {\em smooth} field on $M$---an
$\otimes_{\smooth(M)}$-tensor.}

Of course, the theoretical physicist has time and again proven to
be resourceful and inventive when confronted with such apparently
insurmountable obstacles: for example, in a single stroke she may
throw away the manifold picture of spacetime
altogether\footnote{Albeit, with a heavy heart, since if $M$ has
to go, so will CDG, so the continuous field theory based on it---a
theory which has served her so well in the past: from the ever so
successful (`macroscopically') relativistic field theory of
gravity (GR), to the equally if not more successful
(`microscopically') flat quantum field theories (QFT) of matter.}
and opt for a `discrete', finitistic model of spacetime and
gravity. For, in any case, the general feeling nowadays is that
very strong gravitational fields---probing  smaller and smaller
spacetime scales---such as those developing in the vicinity of a
black hole whose horizon is usually regarded as concealing a
singularity in its core ({\it eg}, the Schwarzschild black hole),
only a quantum theory of gravity will be able to describe
consistently (conceptually) and finitely
(`calculationally').\footnote{Notwithstanding that singularities
are normally regarded as a problem of GR {\it per se} ({\it ie},
of {\em classical} gravity), long before quantization becomes an
issue.} The implicit rationale (or at least the hope) here is that
as the process of quantization ({\it ie}, the development of QFT)
somewhat alleviated the singularities and associated infinities of
the classical field theories of matter ({\it eg}, QED relative to
classical Maxwellian electrodynamics), in the same way a
quantization of GR may heal the singularities and related
pathologies of the classical theory (even though the QFTheoretic
formalism still essentially relies on a background spacetime
continuum, be it flat Minkowski space). There is also an even more
`iconoclastic' stance maintaining that both GR and quantum theory
have to be modified somehow to achieve a fruitful unison of the
two into a consistent QG, which will then be able to shed more
light on, if not resolve completely, the problem of singularities
in GR \cite{pen5}.

In other words, it is generally accepted today that GR appears to
be out of its depth when trying to describe the gravitational
field right at its source ({\it eg}, the inner Schwarzschild
singularity of the gravitational field of a point-particle
\cite{df}), while at the same time, below the so-called Planck
time-length---or equivalently, in dynamical processes
(interactions) of very high energy-momentum transfer where quantum
gravitational effects are supposed to become important---the
classical spacetime continuum of macroscopic physics should give
way to something more `reticular' and `quantal'.

{\it In summa}, on the face of the aforementioned impasse and the
subsequent hopes that QG could (or maybe, {\em should}?) remove
singularities and their associated infinities in the end, {\em
there} goes the spacetime manifold and, inevitably, {\em down}
comes the whole CDG-edifice that is supported by it. The
expression `{\em throw the baby away together with the
bath-water}' is perhaps suitable here, with the ever so valuable
baby representing {\em differential geometric concepts and
constructions}, while the epicurical aqueous `bathing medium'
standing for the `ambient' base manifold which apparently (but
{\em only apparently}, as we will see in the sequel) vitally
supports those CDG concepts and constructions. As a matter of
fact, in the past, QG researchers have gone as far as to claim
that

\begin{quotation}

\noindent ``{\small ...at the Planck-length scale, differential
geometry is simply incompatible with quantum theory...[so that]
one will not be able to use differential geometry in the true
quantum-gravity theory...}'' \cite{ish}

\end{quotation}

On the other hand, there is the recently developed {\em Abstract
Differential Geometry} (ADG) \cite{mall1,mall2,mall4}, a theory
and method of doing differential geometry that does not employ at
all a background geometrical $\smooth$-smooth manifold, while at
the same time it still retains, by using purely
algebraico-categorical (:sheaf-theoretic) means, all the
differential geometric conceptual panoply and technical machinery
of the manifold based CDG. This {\em differential geometric
mechanism} of CDG, ADG has taught us both in theory and in
numerous applications so far, is in essence of a purely algebraic
character and quite independent of a base geometrical continuum,
much in the relational way Leibniz had envisioned that Calculus
should be formulated and practiced \cite{leibniz1}. However, that
`{\em fundamental algebraicity}' is masked by the `geometric
mantle' of the background locally Euclidean space(time) $M$ which
intervenes in our differential geometric calculations ({\it ie},
in our Differential Calculus!) in the guise of the smooth
coordinates of (the points of) $M$ in $\smooth(M)$. Thus, {\em CDG
is a background space(time) dependent conception and method of
differential geometry} that could be coined, in contradistinction
to the base manifoldless Leibnizian ADG, Cartesian-Newtonian
\cite{mall7,mall10,malrap4}.

As a result, the relevance of ADG regarding the dilemma-{\it
cum}-impasse posed above is that one not only is {\em not} forced
to throw away the baby (:the differential mechanism) together with
the bath-water (:the base manifold), but also that one can
exercise that essentially algebraic differential geometric
machinery in the very presence of singularities of any kind,
literally as if the singularities were not there
\cite{malros1,mall3,malros2,malros3,mall9,mall7,mall11,malrap4}.
As it happens, ADG passes through the horns of the aforementioned
dilemma by doing away with one horn ({\it ie}, the base spacetime
manifold) while showing at the same time that the gravitational
field law---which is still algebraico-categorically represented by
a base manifoldless version of the {\em differential} equations of
Einstein---holds over, and by no means breaks down at,
singularities of any sort.\footnote{That is to say, not only in
the presence of the usual, `classical' as it were, singularities
which are built into the smooth coordinates $\smooth(M)$ of the
pointed differential manifold $M$, but also with respect to more
general, far more numerous and `robust' ones, such as the
so-called `{\em spacetime foam dense singularities}' teeming
Rosinger's differential algebras of generalized functions
(non-linear distributions). These are functions that are defined
on finite-dimensional Euclidean and locally Euclidean (manifold)
space(time)s, and include not only the smooth functions in
$\smooth(M)$, but also more general, `smeared out' function(al)s,
such as the linear distributions of Schwartz
\cite{malros1,malros2,mall3,malros3,malrap4}.} Consequently, the
latter are not interpreted as being insuperable obstacles to, let
alone break down points of, `differentiability' as the manifold
`mediated' CDG (and consequently, the GR based on it) has so far
(mis)led us to believe \cite{hawk0,clarke4,malrap4}.

In the present paper we put ADG further to the test by applying it
towards the `resolution' (or better, as we shall see in the
sequel, towards the total evasion or bypass) of the interior
singularity of the spherically symmetric Schwarzschild solution of
the (vacuum) Einstein equations for the gravitational field
surrounding a point-particle of mass $m$. Classically ({\it ie},
from the viewpoint of the manifold based CDG and GR), this
singularity, unlike the exterior one located at the so-called
Schwarzschild black hole horizon-radius distance $r=2m$ from the
point-mass which has proven to be merely a `virtual', so-called
{\em coordinate} one \cite{df}, is thought of as being a `real',
`genuine' singularity as it resists any analytic ($\anal$), smooth
($\smooth$), or even continuous ($\cont$), extension of the
spacetime manifold $M$ past it \cite{df,hawk0,clarke4,malrap4}. In
turn, the {\em differential} field equations of Einstein are
thought of as breaking down at the {\it locus} of the point-mass
in the sense that they are no longer regarded as a valid
description of gravitational dynamics right at the source of the
gravitational field. As noted earlier, the general consensus
nowadays is that only a QG will be able to describe gravitational
dynamics for very strong, divergent from the pointed-continuum
perspective and when treated with the usual analytic means of CDG
(Calculus), gravitational fields near their massive
(energy-momentum) sources. Even more dramatically and drastically,
it is intuited that the said `infinitistic' manifold, by evoking a
minimal, fundamental space-time length-duration
($\ell_{P}$-$t_{P}$), should be replaced by a `granular' and
`quantal' structure which correctly represents the `true'
spacetime geometry in the quantum deep \cite{ash4}.

In glaring contrast to the anticipations and hopes above, in this
paper we will show by using the purely algebraic
(:sheaf-theoretic), manifold- and {\it in extenso} Calculus-free
ADG-theoretic means, that the (vacuum) Einstein equations not only
do not break down in any sense, as {\em differential} equations
proper, in the immediate vicinity of, or even right at, the
Schwarzschild point-particle, but also that they hold both at the
`discrete' and at the `continuous' background space(time) level of
description of gravity. To this end, Sorkin's finitary poset
substitutes of continuous manifolds \cite{sork0}, in their
Gel'fand-dual algebraic representation in terms of `discrete
differential incidence algebras' \cite{rapzap1,rapzap2} and the
finitary spacetime sheaves (finsheaves) thereof
\cite{rap2,malrap1,malrap2}, are used {\it \`a la} ADG to show
that the law of gravity (`originating' from the Schwarzschild
point-mass) holds both at the `reticular-quantal' level of
description of spacetime \cite{malrap3} {\em and} in a (suitably
defined) `classical', `continuum' (inverse) limit of (a projective
system of) the said finsheaves and the finitary differential
triads that the latter comprise \cite{malrap2,malrap3}. We infer
what has been already anticipated numerous times in the past
trilogy \cite{malrap1,malrap2,malrap3} of applications of ADG to a
(f)initary, (c)ausal and (q)uantal (abbreviated `$fcq$') version
of Lorentzian gravity, namely, that {\em ADG allows us to develop
a genuinely background spacetime independent, purely gauge} ({\it
ie}, solely connection based) {\em field theory of gravity}, no
matter whether that base `spacetime' is taken to be a continuum or
a discretum.

{\it Ex altis} viewed, the paper is organized as follows: in the
next section we review some ADG-basics from \cite{mall1,mall2}
that will prove to be useful in the sequel. In section 3 we recall
from the trilogy \cite{malrap1,malrap2,malrap3} the essentials
from the ADG-theoretic approach, via Sorkin's finitary
discretizations of continuous manifolds \cite{sork0}, their
Gel'fand-dual incidence algebraic representation
\cite{rapzap1,rapzap2} and the latter's finsheaf-theoretic picture
\cite{rap2}, to a $fcq$-version of Lorentzian vacuum Einstein
gravity. In section 4 we bring forth from
\cite{pap1,pap2,pap3,pap4,pap5} the key result from the
categorical perspective on ADG, namely, that the category of
differential triads is bicomplete---{\it ie}, closed under both
projective and inductive limits. Having that result in hand, in
the following section (5) we present a direct `static' (or
`stationary'), `spatial' (spacetime-localized) point-resolution of
the interior Schwarzschild singularity and we anticipate an
alternative `temporal' (time-line extended), distributional one
involving the so-called spacetime foam dense singularities from
\cite{malros2,malros3}. However, we leave the technical details of
the latter for the more comprehensive treatment of
$\smooth$-gravitational singularities in \cite{malrap4}. The paper
concludes with a brief discussion on the possibility of developing
a genuinely background independent QG and we compare the
Schwarzschild singularity resolution presented herein with similar
recent efforts in the context of LQG \cite{modesto}, passing at
the same time the baton to \cite{malrap4} for a more thorough
exposition of the potential import of ADG-ideas to current QG
research.

\section{Rudiments of ADG}

We first recall from \cite{mall1,mall2} some key concepts and
structures in ADG that will prove to be useful in what follows.

\paragraph{$\mathbf{K}$-algebraized spaces.} In ADG, we let $X$ be
an in principle {\em arbitrary topological space} on which a {\em
sheaf} $\struc$ of unital, commutative and associative
$\mathbb{K}$-algebras $A$ is erected. The coefficient field
$\mathbb{K}$ of the algebras may be taken to be either $\R$ or
$\com$. We tacitly assume that the {\em constant sheaf}
$\mathbf{K}\equiv\cons$ of $\mathbb{K}$-scalars is canonically
embedded (injected) into $\struc$:
$\mathbf{K}\hookrightarrow\struc$. We say that $X$ is the {\em
base space} (for the localization) of the {\em structure sheaf}
$\struc$ of {\em generalized arithmetics}.\footnote{The terms
`{\em coefficients}' or `{\em coordinates}' will be used
interchangeably with the term `{\em arithmetics}' in the sequel.}
The pair

\begin{equation}\label{eq1}
\diad:=(X,\struc_{X})
\end{equation}

\noindent is called a {\em $\mathbf{K}$-algebraized space}.

\paragraph{Vector sheaves and differential triads.} Technically speaking, by a
{\em vector sheaf} $\modl$ in ADG we mean a {\em locally free
$\struc$-module of finite rank}, that is to say, a sheaf of
$\struc$-modules over $X$ that is locally expressible as a finite
power (a finite Whitney sum) of $\struc$

\begin{equation}\label{eq2}
\modl|_{U}\simeq\struc^{n}|_{U}=(\struc|_{U})^{n}=\struc^{n}(U)=\struc(U)^{n}~(U~\mathrm{open~in}~X)
\end{equation}

\noindent with $\struc^{n}(U)=\struc(U)^{n}:=\Gamma(U,\struc)$ the
local sections of $\struc$.

We also assume that {\em the dual of} $\modl$

\begin{equation}\label{eq3}
\modl^{*}:=\Omg(\equiv\Omg^{1})={\Hom}_{\struc}(\modl ,\struc)
\end{equation}

\noindent is the ADG-theoretic analogue of the sheaf of modules of
smooth $1$-forms in the classical, manifold based theory (CDG). It
must be emphasized here that CDG is `recovered' from ADG when one
assumes $\smooth_{X}$ for structure sheaf $\struc$ of coordinates
in the theory, which in turn means that $X$ is a smooth manifold
(Gel'fand duality and spectral theory).

Now, having defined $\diad$s, $\modl$s and their duals $\Omg$, we
are in a position to define {\em the} fundamental notion in ADG,
that of a {\em differential triad} $\triad$. It is a triplet

\begin{equation}\label{eq4}
\triad :=(\struc_{X} ,\partial ,\Omg^{1}_{X})
\end{equation}

\noindent consisting of a structure sheaf $\struc_{X}$ on some
topological space $X$ ({\it ie}, a $\mathbf{K}$-algebraized space
$\diad$ is built into every $\triad$)\footnote{For typographical
economy, from now on we will omit the base space $X$ as a
subscript to the sheaves involved.} and a $\mathbf{K}${\em -linear
Leibnizian sheaf morphism} $\partial$. That is to say, $\partial$
is a map

\begin{equation}\label{eq5}
\partial :~\struc\mapto\Omg^{1}
\end{equation}

\noindent which is $\mathbf{K}$-linear, and for every two local
sections $p$ and $q$ in $\Gamma(U,\struc)\equiv\struc(U)$ (:the
collection of local sections of $\struc$ over $U\subset X$), the
usual Leibniz rule is observed

\begin{equation}\label{eq6}
\partial(p\cdot q)=p\cdot\partial(q)+q\cdot\partial(p)
\end{equation}

\paragraph{$\struc$-connections.} The basic observation of ADG is that {\em the basic differential
operator $\partial$ in differential geometry is the archetypical
instance of an $\struc$-connection}\footnote{In ADG, the concept
of an algebraic $\struc$-connection is {\em the} fundamental one,
about which the whole theory revolves. $\struc$-connections are
the {\it raison d'\^{e}tre} of ADG
\cite{mall-2,mall1,mall2}.}---albeit, a {\em flat} connection as
we shall see below.\footnote{Moreover, in complete analogy to
$\partial$, one can then iteratively define higher order
prolongations $d^{i}$ ($i\geq 1$) of $\partial\equiv d^{0}$, which
again are $\mathbf{K}$-linear and Leibnizian sheaf morphisms
between $\struc$-modules $\Omg^{i}$ of differential form-like
entities of higher degree $d^{i}:~\Omg^{i}\mapto\Omg^{i+1}$
($\struc\equiv\Omg^{0}$), satisfying at the same time the usual
nilpotency condition of the standard (exterior Cartan-de
Rham-K\"ahler) differential operator $d$: $d^{i+1}\circ
d^{i}\equiv d^{2}=0$ (with $d^{2}$ being `{\em the square of
$d$}', not to be confused with the second order prolongation of
$\partial$).} Thus, a general (curved) $\struc$-connection $\conn$
in ADG is an abstraction from and a generalization of the usual
$\partial$, defined as the following $\mathbf{K}$-linear sheaf
morphism

\begin{equation}\label{eq7}
\conn :~\modl\mapto\Omg(\modl)\equiv\modl\otimes_{\struc}\Omg\cong
\Omg\otimes_{\struc}\modl
\end{equation}

\paragraph{Curvature of $\struc$-connections.} With $\conn$ at
our disposal, we can define its curvature $\curv(\conn)$
diagrammatically as follows

\begin{equation}\label{eq8}
\Vtriangle[\modl`\Omg^{1}(\modl)\equiv\modl\otimes_{\struc}\Omg^{1}`\Omg^{2}(\modl)\equiv\modl\otimes_{\struc}
\Omg^{2};\conn`\curv\equiv\conn^{1}\circ\conn`\conn^{1}]
\end{equation}

\noindent for a suitable higher order extension $\conn^{2}$ of
$\conn(\equiv\conn^{1})$.\footnote{Like the higher order
extensions $d^{i}$ of $\partial\equiv d^{1}$ mentioned in the last
footnote, $\conn^{2}$ for example is a $\mathbf{K}$-linear,
Leibnizian sheaf morphism between $\Omg^{1}$ and $\Omg^{2}$:
$\conn^{2}:~\Omg^{1}(\modl)\mapto\Omg^{2}(\modl)$. It acts locally
({\it ie}, section-wise), and relative to $\conn$, as follows:
$\conn^{2}(p\otimes q):=p\otimes d q-q\wee\conn
q,~(p\in\modl(U),q\in\Omg^{1}(U),U~\mathrm{open~in}~X)$.} It must
be noted here that, from the definition of $\curv(\conn)$ above,
it follows that the nilpotent $\partial$ is a {\em flat}
connection---{\it ie}, $\curv(\partial):=d^{2}=0$. It is also
important to observe here that, unlike $\conn$ which is only a
$\mathbf{K}$-sheaf morphism, its curvature $\curv(\conn)$ is an
$\struc$-morphism, that is to say, our generalized arithmetics
(coordinates) in $\struc$ respect it. Equivalently, and
philologically speaking, $\curv$ `sees through' our generalized
arithmetics (coordinates) in $\struc$. On the other hand, our acts
of coordinatization in $\struc$ cannot `capture' $\conn$, which
eludes them since it is not an $\struc$-morphism
\cite{mall1,malrap2,malrap3}.\footnote{This observation about
$\curv(\conn)$ will become important when we discuss the {\em
$\struc$-functoriality} of the ADG-theoretic formulation of
(vacuum) gravitational dynamics in (\ref{eq9}) below.}

\paragraph{Manifoldless (pseudo-)Riemannian geometry and vacuum Einstein equations in a nutshell.}
Following \cite{mall1,mall2,mall3,malrap1,malrap2,malrap3}, we can
then formulate ADG-theoretically, {\em in a manifestly background
(spacetime) manifoldless way}, all the concepts and structures of
the CDG-based (pseudo-)Riemannian geometry underlying GR such as
$\struc$-valued (Lorentzian) metrics $\rho$, Christoffel
$\struc$-connections $\nabla$ compatible with $\rho$ ({\it ie},
metric or torsionless connections), the Ricci curvature $\modl
nd\modl$-valued\footnote{And recall that, locally: $\modl
nd\modl(U)=M_{n}(\struc)(U)$.} $\otimes_{\struc}$-tensor $\ric$,
and its $\struc$-valued trace-contraction---the Ricci scalar
$\ricci$.

The upshot of our brief {\it r\'esum\'e} here of the application
of ADG to GR is that the vacuum Einstein equations read in our
scheme

\begin{equation}\label{eq9}
\ricci(\modl)=0
\end{equation}

\noindent recalling at the same time from
\cite{mall3,malrap3,malrap4} a couple of important observations
pertaining to them:

\begin{enumerate}

\item {\em From the ADG-theoretic viewpoint, GR is another gauge
theory}---in fact, a `{\em pure gauge theory}' as the only
dynamical variable involved is the (curvature of the)
gravitational $\struc$-connection $\conn$, and no external smooth
base spacetime manifold is employed. This is in glaring contrast
to {\em both} the original (smooth) spacetime metric-based
formulation of GR by Einstein (2nd-order formalism) {\em and} to
the recent `new variables' formulation of GR by Ashtekar
\cite{ash} (1st-order formalism reminiscent of Palatini's
metric-affine one) which, although it emphasizes the importance of
the notion of connection so as to place gravity in the category of
gauge forces, it still employs a smooth background manifold, while
at the same time the (smooth) metric of the 2nd-order formalism is
still present `in disguise', being encoded in the (smooth) {\it
vierbein} field variables. Due to these features, we coin the
ADG-formulation of GR `{\em half-order, pure gauge
formalism}'.\footnote{`{\em Half-order}', because only $\conn$,
and not $g_{\mu\nu}$ (2nd-order) or $e_{\mu}^{a}$ (1st-order), is
engaged in the dynamics (and in the 1st-order formalism there are
{\em two} basic variables engaged in the dynamics: the
$\smooth$-connections and the smooth comoving frame-tetrads).
`{\em Pure gauge}', because there is no `external' spacetime
(manifold) involved---only `internal', gauge `degrees of freedom'
associated with the gravitational connection field $\conn$
`in-itself'. In the concluding section we will return to comment
further on this virtue of the ADG-formulation of gravity and its
implications for developing a genuinely background independent
QG.}

\item The vacuum Einstein equations derive variationally (solely with
respect to $\conn$!) from the ADG-theoretic version of the
Einstein-Hilbert action functional $\eh$, which is an
$\struc$-valued functional on the affine space
$\sconn_{\struc}(\modl)$ of the $\struc$-connections $\conn$,
which in turn becomes the relevant configuration space in our
theory of gravity.

\item Since there is no external smooth spacetime continuum involved in
the ADG-version of GR, the principle of general covariance (PGC)
of the usual manifold based theory is not expressed via
$\mathrm{Diff}(M)$ as usual, but via $\aut\modl$---the (group
sheaf of) automorphisms (dynamical self-transmutations) of the
gravitational field itself. Here one might wish to recall that in
ADG the term {\em field} pertains to the pair $(\modl ,\conn)$,
with $\modl$ the (geometric) representation (or carrier) space of
the (algebraic) connection field $\conn$
\cite{mall1,mall2,malrap3,malrap4}. In technical jargon, $\modl$
is the associated (representation) sheaf of the principal sheaf
$\aut\modl$ of field automorphisms \cite{vas1,vas2,vas3,vas4}.
Moreover, since $\modl$ is by definition locally isomorphic to
$\struc^{n}$, $\aut\modl(U):=(\modl
nd\modl(U))^{\bull}\equiv(M_{n}(\struc))^{\bull}$. This is an
autonomous conception of covariance, pertaining directly to the
gravitational field `in-itself', without reference to an external
spacetime manifold, which we have elsewhere coined  `{\em
synvariance}'.

In connection with our remarks above about gravity as a `pure
gauge theory' {\it \`a la} ADG, no {\em external} spacetime
(manifold) symmetries in the guise of $\mathrm{Diff}(M)$ appear in
our theory---only the `{\em internal}', gauge ones $\aut\modl$ of
the field $(\modl ,\conn)$ `in-itself' are involved. In fact, the
distinction external/internal symmetries loses its meaning in our
ADG-perspective on gravity. Of course, assuming $\smooth_{X}$ for
structure sheaf---{\it ie}, that $X$ is a differential manifold
$M$---one may recover, if one wishes, the external
$\mathrm{Diff}(M)$ used in the mathematical expression of the PGC
of the CDG and smooth manifold based GR since, by definition,
$\mathrm{Aut}M\equiv\mathrm{Diff}(M)$. It also follows now that
the relevant configuration space is the aforementioned affine
space $\sconn_{\struc}(\modl)$ of $\struc$-connections modulo the
field's dynamical self-transmutations (`autosymmetries') in
$\aut\modl$: $\sconn_{\struc}(\modl)/\aut\modl$.

\item Finally, closely related to the remarks about synvariance above is the issue
of {\em functoriality}. In the ADG perspective on GR,
functoriality pertains to the fact that the gravitational
dynamics---the vacuum Einstein equations (\ref{eq9})---is
expressed via the curvature of the connection, which is an
$\struc$-morphism---or equivalently, a $\otimes_{\struc}$-tensor
($\otimes_{\struc}$ being the homological tensor product functor).
This means that {\em the generalized coordinates in $\struc$, that
{\em we} employ in order to `measure' or `geometrically represent'
(and `localize' in $\modl$ over $X$) the gravitational connection
field $\conn$, respect it}.\footnote{Although it must be stressed
here that the connection itself, being simply a
$\mathbf{K}$-morphism, is {\em not} an $\struc$-morphism or
$\otimes_{\struc}$-tensor, thus it `eludes' our measurements in
$\struc$. However, it is the curvature of the connection that
appears in (\ref{eq9}), which {\em is} an $\struc$-morphism.
$\conn$ is a purely algebraic notion, and as such it evades our
generalized acts of measurement or `geometrization' (and
concomitant representation on the associated sheaf $\modl$) of the
gravitational field $\conn$, which are organized in $\struc$
\cite{malrap3,malrap4}.} Moreover, since if any space(time) is
involved at all in our scheme, then it is regarded as being built
into the $\struc$ that {\em we} assume in the first place to
coordinatize (or geometrically represent) the gravitational
connection field $\conn$ (on $\modl$),\footnote{What we have in
mind here is a generalized version of the notion of Gel'fand
duality whereby, in the same way that in the classical theory
(CDG) one obtains a smooth manifold $M$ as the Gel'fand spectrum
of topological algebra $\smooth(M)$ (or equivalently, from
$\struc\equiv\smooth_{M}$) \cite{mall0,mall-1}, one can
(spectrally) extract other `geometrical' base space(time)s from
various different choices of structure algebra sheaves $\struc$
(indeed, by assuming `functional' structure sheaves other than
$\smooth_{M}$).} the gravitational dynamics, being
$\struc$-functorial, `sees through' the said `{\em spectral
space(time)}' inherent in $\struc$.

Precisely in this $\struc$-functoriality lies the strength and
import of ADG {\it vis-\`a-vis} (gravitational) singularities, in
the sense that one can `{\em absorb}', {\em incorporate}, or {\em
integrate} into $\struc$ singularities of any kind---ones that are
arguably more robust and numerous than the $\smooth$-ones built
into the usual coordinate structure sheaf $\smooth_{M}$ of the
smooth manifold---and still be able to show that the gravitational
field equations hold and in no way break down at their {\it loci}
in $X$. As it were, {\em the differential equations of Einstein
hold over and above them, in spite of their presence in the
$\struc$ being employed}
\cite{malros1,malros2,mall3,mall9,malros3,mall11,malrap4}.

\end{enumerate}

\paragraph{The categorical imperative of ADG.} Throughout the present paper we have mentioned various
category-theoretic sounding terms, as for example the notions of
{\em sheaf morphism} and {\em functoriality}. Indeed, on the whole
one can say that ADG is an algebraico-categorical scheme for doing
differential geometry \cite{mall1,mall2}, for after all, ``{\em
the methods of sheaf theory are algebraic}'' \cite{grarem}. Here
we expose briefly some key categorical aspects of ADG as explored
in great depth in \cite{pap1,pap2,pap3,pap4,pap5}.

The first thing to mention is that one can regard differential
triads as objects in a category $\ctriad$---{\em the category of
differential triads} \cite{pap1,pap5}. The arrows in $\ctriad$ are
{\em triad morphisms}, whose definition we now readily recall from
\cite{pap1,pap4,pap5}

One lets $X$ and $Y$ be topological spaces, assumed to be the base
spaces of some $\mathbf{K}$-algebraized spaces $(X,\struc_{X})$
and $(Y,\struc_{Y})$, respectively. In addition, one lets
$\triad_{X}=(\struc_{X},\partial_{X},\Omg_{X})$ and
$\triad_{Y}=(\struc_{Y},\partial_{Y},\Omg_{Y})$ be differential
triads over them. Then, a {\em morphism $\morph$ between
$\triad_{X}$ and $\triad_{Y}$ is a triplet of maps}
$\morph=(f,f_{A},f_{\Omg})$, enjoying the following four
properties:

\begin{enumerate}

\item the map $f:~X\mapto Y$ is {\em continuous};

\item the map $f_{\struc}:~\struc_{Y}\mapto f_{*}(\struc_{X})$ is
a {\em morphism of sheaves of} $\mathbb{K}${\em -algebras over}
$Y$ preserving the respective algebras' unit elements ({\it ie},
$f_{\struc}(1)=1$);\footnote{In the expression for $f_{\struc}$
above, $f_{*}$ is the {\em push-out} along the continuous $f$, a
map which carries each element of a differential triad into a like
element in the sense that, for any triad $\triad$,
$f_{*}(\triad):=(f_{*}(\struc),f_{*}(\partial),f_{*}(\Omg))$ is
also a differential triad---the one `induced' by $f$
\cite{pap4,pap3}; whence, term-wise for our triads $\triad_{X}$
and $\triad_{Y}$ above (and omitting the base topological space
subscripts):
$f_{*}(\struc):=(f_{*}(\struc)(U):=\struc_{f^{-1}(U)}),~(U\subseteq
Y~\mathrm{open})$ is a sheaf of unital, abelian, associative
$\mathbb{K}$-algebras over $Y$,
$f_{*}(\Omg):=(f_{*}(\Omg)(U):=\Omg_{f^{-1}(U)}),~(U\subseteq
Y~\mathrm{open})$ a sheaf of $f_{*}(\struc)$-modules (of 1st-order
differential form-like entities), and
$f_{*}(\partial):=(f_{*}(\partial)(U):=\partial_{f^{-1}(U)}),~(U\subseteq
Y~\mathrm{open})$ an induced $\mathbf{K}$-linear, Leibnizian sheaf
morphism \cite{pap4}.} and the following categorical diagram is
obeyed:

\[
\bfig
\putsquare<1`-1`-1`1;700`700>(700,700)[X`Y`\struc_{X}`\struc_{Y};f```]
\putbtriangle<0`-1`0;700>(1400,700)[``~~~f_{*}(\struc_{X});`f_{\struc}`]
\putmorphism(1400,700)(1,0)[\phantom{TT'T'}`%
\phantom{TT'}`]{700}1a \efig
\]

\item the map $f_{\Omg}:~\Omg_{Y}\mapto f_{*}(\Omg_{X})$, as noted
in the last footnote, is a {\em morphism of sheaves of}
$\mathbb{K}${\em -vector spaces over} $Y$, with
$f_{\Omg}(\alpha\omega)=f_{\struc}(\alpha)f_{\Omg}(\omega),~\forall
\alpha\in\struc_{Y},~\omega\in\Omg_{Y}$; and finally,

\item with respect to the $\cons\equiv\mathbf{K}$-sheaf morphism ({\it viz.} flat
connection) $\partial$ in the respective triads, and as it has
also been alluded to in the last footnote, the following diagram
is commutative:

\[
\bfig
\putsquare<1`1`1`1;600`600>(600,600)[\struc_{Y}`\Omg_{Y}`f_{*}(\struc_{X})`f_{*}(\Omg_{X});
\partial_{Y}`f_{\struc}`f_{\Omg}`f_{*}(\partial_{X})]
\efig
\]

\noindent which reads: $f_{\Omg}\circ
\partial_{Y}=f_{*}(\partial_{X})\circ f_{\struc}$.

\end{enumerate}

\noindent {\it In summa}, {\em $\ctriad$ is a category having
$\triad$s for objects and $\morph$s for arrows}. Let it be noted
here that in the past it has been amply observed that differential
triads are generalizations of differential manifolds. Indeed, the
entire differential structure of a $\smooth$-smooth manifold $M$
is encoded in the {\em classical differential triad}
$\triad_{\infty}$ having as $\struc$ the sheaf of germs of local
($\mathbb{K}\equiv\R$-valued) $\smooth$-functions on $M$, as
$\Omg$ the usual sheaf of germs of local $\smooth$-differential
$1$-forms ({\it ie}, $\Omg\equiv\Gamma^{\infty}(T^{*}M)$), and one
can identify $\partial$ with the usual (exterior) derivative $d$:
$\partial\equiv d:~\struc\mapto\Omg
:~\alpha\in\struc\mapsto\partial(\alpha):=d\alpha\in\Omg$. It must
be also stressed that $\triad_{\infty}$ is only a particular
instance of the general (abstract notion of) differential triad,
which, as noted earlier, is able to accommodate algebraized spaces
(and differentials $\partial$ on them) other than the classical
one $\diad_{\infty}=(M,\smooth_{M})$ (and $\partial\equiv
d$)---{\it ie}, algebraized spaces hosting structure sheaves other
than $\smooth_{X}$, and possibly non-functional (of course, as
long as these generalized arithmetics provide one with the fertile
ground on which to define a $\partial$ or $\conn$ {\it \`a la}
(\ref{eq5}) or (\ref{eq7}), and thus to develop differential
geometric ideas with them).

But let us discuss a bit more this categorical versatility of the
differential triads of ADG compared to the `rigidity' and
associated shortcomings of (the category of) smooth manifolds.

\paragraph{Brief discussion of the categorical versatility of
ADG.} The categorical `versatility' and `flexibility' of ADG,
compared to the `crystalline rigidity' of the manifold based CDG,
may be summarized by outlining the following shortcomings of
$\man$---the category of (finite dimensional) differential
($\smooth$-smooth) manifolds---relative to $\ctriad$:

\begin{enumerate}

\item {\em $\man$ has no initial or final structures. That is, one cannot pull-back
or push-out a smooth atlas by a continuous map.}

\item {\em The quotient space of a manifold by an (arbitrary) equivalence
relation is not a manifold}.

\item {\em Similarly, an arbitrary subset of a manifold is not a
manifold. In other words, $\man$ has no canonical subobjects.}

\item {\em In general, $\man$ is not closed under inductive (direct)
or projective (inverse) limits.\footnote{In category-theoretic
jargon, projective (inverse) limits are known as `{\em categorical
limits}', while inductive (direct) ones as `{\em categorical
colimits}'.} Another way to say this is that $\man$ is not
bicomplete ({\it ie}, complete and co-complete)}.

\item {\em Generally, there are no well defined categorical products
and co-products in $\man$}.

\end{enumerate}

\noindent As Papatriantafillou has shown in a long series of
thorough investigations \cite{pap1,pap2,pap3,pap4,pap5}, $\ctriad$
not only does not suffer from such (differential geometric)
maladies, but also goes all the way towards healing or bypassing
them completely. Thus, from a mathematical point of view alone,
theoretical physics' applications aside, these differential
geometric anomalies of $\man$ could suffice for motivating the
development of ADG---in fact, they could be regarded as the {\it
raison d'\^{e}tre et de faire} of ADG. In particular, and of
special importance to the present paper as we shall see in the
sequel, Papatriantafillou has shown in connection with the
differential manifolds' deficiencies 1, 2 and 4 above, that in
$\ctriad$:

\begin{itemize}

\item And we quote, ``{\em the differential mechanism induced by a differential
triad is transferred backwards and forward by any continuous map
$f$. The initial and final structures thus obtained satisfy
appropriate universal conditions that turn the continuous map $f$
into a differentiable map.}'' \cite{pap3,pap4}. To recapitulate in
a nutshell this result, given a continuous map $f:~X\mapto Y$,
with $X$ the base space of a differential triad $\triad_{X}$,
Papatriantafillou showed that $f$ pushes forward the (essentially
algebraic) differential mechanism of $\triad_{X}$, so that a new
and unique differential triad---one that satisfies a {\em
universal mapping} condition \cite{pap4}---is defined on $Y$, so
that in the process, {\em $f$ becomes differentiable}. The
relevant theorem,\footnote{Theorem {\bf 3.1} in \cite{pap3}.}
which uses some ideas already mentioned {\it en passant} in
footnote 18 before, can be stated as follows:\footnote{For the
corresponding detailed proof, the reader is referred to
\cite{pap3}.}

\ovalbox{\bf Theorem:} Let
$\triad_{X}=(\struc_{X},\partial_{X},\Omg_{X})~\in\ctriad_{X}$,\footnote{Plainly,
$\ctriad_{X}$ is the subcategory of $\ctriad$ consisting of all
differential triads and triad morphisms with common base
topological space $X$.} and $f:~X\mapto Y$ continuous. When $Y$
inherits
$f_{*}(\triad_{X}):=(f_{*}(\struc_{X}),f_{*}(\partial_{X}),f_{*}(\Omg_{X}))$
from the push-out $f_{*}$ of $f$, then there is a morphism of
differential triads
$\morph=(f,f_{\struc},f_{\Omg}):~\triad_{X}\mapto
f_{*}(\triad_{X})~ (\in\ctriad)$---{\it ie}, $f$ becomes {\em
differentiable}. Moreover, the pushed-forward triad
$f_{*}(\triad_{X})$ satisfies the following universal
(composition) property \cite{pap4}: given a triad
$\triad_{Y}=(\struc_{Y},\partial_{Y},\Omg_{Y})\in\ctriad_{Y}$, and
a morphism
$\tilde{\morph}:=(f,\tilde{f}_{\struc},\tilde{f}_{\Omg}):~\triad_{X}\mapto\triad_{Y}$,
there is a unique morphism
$(id_{Y},g_{\struc},g_{\Omg}):~f_{*}(\triad_{X})\mapto\triad_{Y}$
such that
$\tilde{\morph}=(id_{Y},g_{\struc},g_{\Omg})\circ\morph$.

Accordingly, the `dual' (converse) scenario involving $f$'s {\em
pull-back action} $f^{*}$, when now the range of $f$ is a
differential triad $\triad_{Y}$ on $Y$ while $X$ ($f$'s domain) is
simply a topological space not being endowed {\it a priori} with a
differential (triad) structure, $f^{*}$ too can be seen to
transfer (induce) on $X$ the differential mechanism encoded in
$\triad_{Y}$, thus rendering $X$ a {\em differential} (not just a
topological) space and in the process promoting $f$ to a {\em
differentiable} (not just a continuous) map \cite{pap3}.

Of great mathematical interest is that these results may serve as
the starting point for research into what one might call the `{\em
differential geometry of topological spaces}', and they depict
some sort of `Calculus-reversal', since in the usual theory, `{\em
differentiability implies continuity}', while here in some sense
`{\em continuity} ({\it ie}, topology, plus algebraic
structure---{\it eg}, the employment of a topological vector space
structure) {\em entails differentiability}'. Indeed, {\em
differentiability} ({\it ie}, the ability to define a
derivative/differential operator) {\em is a topologico-algebraic
notion}---one that is secured in the manifold based CDG exactly
because $\smooth(M)$ is a (non-normable) topological algebra
\cite{malrap3,malrap4}.

\item When a manifold $M$ is factored by an equivalence relation
$\sim$, and there happens to be a continuous map $f$ from $M$ to
the resulting quotient space $\tilde{M}=M/\sim$ (suitably
topologized), then the result in 1 above secures that the
classical differential structure ({\it ie}, differential triad) on
$M$ can be pushed-forward by $f_{*}$ on the `moduli space'
$\tilde{M}$, thus endow it with a differential triad of its own.
In the next section we will encounter a concrete example of this
`{\em differential triad induction from a continuum to a
discretum}' having to do with Sorkin's finitary $T_{0}$-poset
discretizations of continuous ($\cont$) manifold topologies
\cite{sork0}.

\item Finally, as Papatriantafillou has shown in \cite{pap2} and in the forthcoming
monograph \cite{pap5}, $\ctriad$, unlike $\man$, is
bicomplete---that is to say, it is closed under projective and
inductive limits. This virtue of $\ctriad$ will prove to be of
paramount importance on the one hand in section 4, where we give
the `classical continuum limit' of $fcq$-differential triads and
of the $fcq$-version of the vacuum Einstein equations (\ref{eq9})
holding on them, and on the other, in section 5, where we provide
an explicit, `constructive' evasion of the interior Schwarzschild
singularity by finitistic-algebraic means as already developed
under the prism of ADG (and briefly summarized in the next
section) in the past tetralogy
\cite{malrap1,malrap2,malrap3,malrap4}.

\end{itemize}

\section{Application of ADG to Finitary, Causal and Quantal Lorentzian Gravity}

For expository completeness, let us first recall from the trilogy
\cite{malrap1,malrap2,malrap3} the basic results and constructions
that led us to formulate an $fcq$-version of vacuum
Einstein-Lorentzian gravity with the help of ADG as these will be
used in section 5 to achieve our main goal here, namely, to evade
the inner Schwarzschild singularity purely finitistically and
algebraically, and in a `constructive' fashion.

\paragraph{Sorkin's finitary substitutes of continuous manifolds: topology (`continuity') from order.}
A brief history of $fcq$-vacuum Einstein gravity begins with
Sorkin's finitary poset discretizations of continuous ({\it ie},
topological, otherwise known as $\cont$-) manifolds.

The original idea in \cite{sork0} is, starting with an open
bounded region $X$\footnote{By `bounded' it is meant that $X$'s
closure is compact, a space otherwise known as {\em relatively
compact}.} in a manifold $M$\footnote{Let it be stressed here that
Sorkin was interested only in the standard continuous ($\cont$-)
topology of $M$ and no allusion to its differential (smooth)
structure was made. Also, there is no harm in assuming the usual
Hausdorff ($T_{2}$) topology for $M$, although Sorkin's results
follow even from a weaker $T_{1}$ assumption.}, to cover it with a
locally finite open covering $\gauge_{i}$. One may recall that a
cover $gauge_{i}$ of $X$ is called locally finite whenever every
point of $X$ has an open neighborhood about it that meets a finite
number of the covering sets. The index `$i$' of the open covering
will be explained shortly. Then, it was observed that $X$ can be
replaced by a `discrete'\footnote{From now one we will often put
`discrete' in single quotation marks so that one does not confuse
it with the technical term `discrete topological space' referring
to the (trivially Hausdorff) topology of a totally disconnected
set, all the points of which are `clopen' ({\it ie}, closed and
open). Even when these quotation marks are omitted, we do {\em
not} mean the set with completely disjoint points, unless
specifically noted.} $T_{0}$-topological space $P_{i}$, having the
structure of a poset, when the following equivalence binary
relation $\sim$ relative to $\gauge_{i}$ is imposed on its points:

\begin{equation}\label{eq12}
\begin{array}{c}
\forall x,y\in
X:~x\stackrel{\gauge_{i}}{\sim}y\Leftrightarrow\Lambda(x)|_{\gauge_{i}}=\Lambda(y)|_{\gauge_{i}}\cr
\Lambda(x)|_{\gauge_{i}}:=\bigcap\{ U\in\gauge_{i}:~x\in U\}
\end{array}
\end{equation}

\noindent where, clearly, $\Lambda(x)|_{\gauge_{i}}$ is the
`smallest' open set in $\gauge_{i}$ containing $x$, which we here
coin `{\em Sorkin's {\it ur}-cell}' (relative to $\gauge_{i}$).

The aforementioned $T_{0}$-poset $P_{i}$, called `the finitary
substitute of the continuous topology of $X$', is then obtained as
the quotient of $X$ by $\stackrel{\gauge_{i}}{\sim}$:

\begin{equation}\label{eq13}
P_{i}=X/\stackrel{\gauge_{i}}{\sim}
\end{equation}

\noindent Plainly, the elements of $P_{i}$ are
$\stackrel{\gauge_{i}}{\sim}$-equivalence classes of $X$'s points,
with the equivalence relation being interpreted as
`indistinguishability' or `non-separability' of $X$'s points by
the covering sets of $\gauge_{i}$. In other words, the `points' of
$P_{i}$ are Sorkin's {\it ur}-cells $\Lambda(x)|_{\gauge_{i}}$
while the points of the original space(time) $X$ have been
substituted, `blown up', or even `smeared' so to speak, by
`larger' open sets about them. Sorkin initially appreciated that
operationally realistic determinations (`measurements') of
space(time) locution can be modelled after coarse regions in the
said space(time), while the continuum, the `sharp' points of which
``{\em carrying its continuous topology}'' \cite{sork0}, is an
ideal theoretical construct not corresponding to ``{\em what we
actually do to produce spacetime by our measurements}''
\cite{sork1}. Let us note {\it en passant}, the said `operational
pragmatism' aside, that it is widely recognized today that the
pathologies of the spacetime continuum ({\it eg}, the
singularities of GR, or even the unphysical infinities of QFT) are
mainly due to its ideal, point-like character, or equivalently, of
the ideal point-like `nature' of the matter sources (:particles)
of the fields involved. Arguably, quantum (field) theory goes some
way towards alleviating the infinities assailing its classical
counterpart exactly because it sets a fundamental limit (a
regularization cut-off scale of the order of Planck) to the ideal
assumption in the classical (field) theory of infinite (spacetime)
point-localization of the relevant fields, which in turn in the
quantum theory are usually modelled after `smeared' and `blown-up'
(operator-valued) distributions.

The important interpretation of the $P_{i}$s in \cite{sork0} as
discrete approximations of the topological manifold $X$ comes from
considering an {\em inverse system} (or net) $\inv=\{ P_{i}\}$ of
such finitary substitutes, and of continuous surjection maps
$f_{ji}$ between them, in the sense that

\begin{equation}\label{eq14}
P_{i}\preceq P_{j}\Leftrightarrow
P_{j}\stackrel{f_{ji}}{\mapto}P_{i}
\end{equation}

\noindent where $\preceq$ {\em is the act of topological
refinement of the} $P_{i}$s\footnote{Roughly, the partial order
$P_{i}\preceq P_{j}$, which comes from the partial ordering of the
$\gauge_{i}$s in an $i$-indexing net thereof and reading `{\em the
open covering $\gauge_{j}$ is finer than $\gauge_{i}$}' (or
equivalently, that the subtopology $\tau_{i}$ of $X$ generated by
finite intersections of arbitrary unions of the $U$s in
$\gauge_{i}$ is included in the corresponding $\tau_{j}$:
$\tau_{i}\subseteq\tau_{j}$---{\it alias}, $\tau_{i}$ is coarser
than $\tau_{j}$), means that there is a continuous surjection
$f_{ji}$ from the topological $T_{0}$-poset $P_{j}$ to $P_{i}$.
The epithet `continuous' for $f_{ji}$ pertains to the fact that
one can assign a `natural' topology---the so-called Sorkin
lower-set topology---to the $P_{i}$s, whereby an open set is of
the form $\mathcal{O}(x):=\{ y\in P_{i}:~y\mapto x\}$, and where
$\mapto$ is the partial order relation in $P_{i}$ (with basic open
sets involving the links or covering---`immediate
arrow'---relations in $P_{i}$). Plainly then, $f_{ji}$ is a
monotone (partial order-preserving) surjection from $P_{j}$ to
$P_{i}$, hence continuous with respect to the Sorkin topology.}
corresponding to the employment of more numerous and `smaller'
open sets ({\it ie}, finer-and-finer $\gauge_{i}$s) to cover $X$'s
points.

Now, the central result in \cite{sork0}, one that qualifies the
$P_{i}$s as genuine topological approximations of the continuum
$X$, is that {\em the said inverse (projective) system $\inv$
possesses an inverse (projective) limit space---call it
$P_{\infty}=\underset{\infty\leftarrow i}{\lim}{\inv}$---that is
practically homeomorphic to the original $\cont$-manifold that we
started with}.\footnote{The adverb `practically' above pertains to
the result from \cite{sork0} that, at the inverse limit of $\inv$,
one does {\em not} actually recover the topological manifold $X$
itself, but a non-Hausdorff space $P_{\infty}$ which includes $X$
as a dense subset. However, one can get back $X$ from
$P_{\infty}$, by a procedure commonly known as {\em Hausdorff
reflection}, as the set of the latter's closed points
\cite{kopperman}.} The physical interpretation of the inverse
limit procedure is that {\em as one employs finer-and-finer open
sets to cover $X$'s points, at the limit of infinite refinement of
the corresponding $\gauge_{i}$s, one obtains a space that is
essentially topologically indistinguishable from (or topologically
equivalent---{\it ie}, homeomorphic---to) the original continuum
$X$}.

It must be also stressed here that in \cite{sork0} a key attribute
of the $P_{i}$s---one that enables one set up the projective
system $\inv$ in the first place and then take its inverse
limit---is that continuous surjections $f_{i}$, corresponding to
`canonical' projection maps from $X$ to the
$\stackrel{\gauge_{i}}{\sim}$-moduli spaces $P_{i}$ \cite{sork0},
enjoy a {\em universal mapping property} which can be expressed by
the diagram below:

\begin{equation}\label{eq15}
\qtriangle[X`P_{j}`P_{i};f_{j}`f_{i}`f_{ji}]
\end{equation}

\noindent That is, $f_{i}=f_{ji}\circ f_{j}$, and it reads that
{\em the map (canonical projection) of $X$ onto the finitary
substitutes is universal among maps into $T_{0}$-spaces}, with
$f_{ji}$ the {\em unique} map---itself an order-monotone
surjection of $P_{i}$ onto $P_{j}$\footnote{Which, as noted
earlier, corresponds to the act of topological coarse-graining
$\gauge_{i}\preceq\gauge_{j}$ ($i\leq j$ in some `refinement
index-net').}---mediating between the continuous projections
$f_{i}$ and $f_{j}$ of $X$ onto the $T_{0}$-posets $P_{i}$ and
$P_{j}$, respectively. With these `canonical' continuous
projections of $X$ onto the $P_{i}$s, the said inverse system of
finitary posets can be written as a collection of triplets $\inv
:=\{(P_{i}, f_{i}, f_{ji})\}$; while formally, the inverse limit
result above can now be cast as
$P_{\infty}=\underset{\infty\leftarrow
j}{\lim}f_{ji}(P_{i})\stackrel{\mathrm{homeo.}}{\simeq}X$ (modulo
Hausdorff reflection). This universal mapping property of the maps
between the finitary $T_{0}$-posets is completely analogous to the
one possessed by the differential triad morphisms (push-outs and
pull-backs) mentioned earlier. In fact, in the paragraph after the
next, when we will discuss finitary differential triads and their
inverse limits, the ideas of Sorkin and Papatriantafillou will
appear to be tailor-cut for each other; albeit, with the ADG-based
work of Papatriantafillou adding an important {\em differential}
geometric twist to Sorkin's originally purely topological ideas.

\paragraph{Incidence algebras of finitary posets: differential structure (`smoothness') from
algebra.} In a pair of papers in collaboration with Zapatrin
\cite{rapzap1,rapzap2}, a so-called {\em incidence Rota algebra}
$\omg_{i}$ was associated, by {\em Gel'fand duality}, with every
$P_{i}$. One formally writes the correspondence as:

\begin{equation}\label{eq16}
P_{i}\mapto\omg_{i}(P_{i})
\end{equation}

\noindent The $\omg_{i}$s\footnote{From now on we drop the
$(P_{i})$ arguments from the $\omg_{i}$s.} were seen to be unital,
associative, but in general non-commutative,\footnote{They are
abelian when the $P_{i}$s are discrete ({\it ie}, completely
disconnected, trivially Hausdorff) topological spaces.}
$\mathbb{K}$-algebras, which {\it a fortiori} are {\em
$\mathbb{Z}_{+}$-graded discrete differential algebras}
(manifolds)

\begin{equation}\label{eq17}
\omg_{i}= \bigoplus_{n\in\Z_{+}}\omg_{i}^{n}=
 \stackrel{\aconn_{i}}{\overbrace{\omg_{i}^0}}\oplus \stackrel{\mathcal{R}_{i}}{\overbrace{\omg_{i}^1
 \oplus\omg_{i}^{2}\oplus\ldots}}\equiv\aconn_{i}\oplus\mathcal{R}_{i}
\end{equation}

\noindent with $\aconn_{i}$ an abelian subalgebra of
$\omg_{i}$\footnote{$\aconn_{i}$, generated by the
`self-incidences' ({\it ie}, the reflexive relations of the
points) in the underlying poset $P_{i}$, is a discrete analogue of
the algebra $\smooth(M)$ of coordinates (or of points, by Gel'fand
duality/spectral theory) on a smooth manifold $M$.} and
$\mathcal{R}_{i}$ a graded differential
$\aconn_{i}$-module.\footnote{$\mathcal{R}_{i}$ is a discrete
analogue of the classical $\smooth(M)$-module of smooth
differential forms on a differential manifold $M$. Each
$\omg_{i}^{n}$ in $\mathcal{R}_{i}$ is a linear subspace of
$\omg_{i}$.} Indeed, there is a discrete version $d_{i}$ of the
usual nilpotent Cartan-de Rham-K\"ahler differential operator
effecting $\mathbb{K}$-linear grade-raising transitions of the
sort $d_{i}:~\omg^{n}\mapto\omg^{n+1}$.

The careful reader will have perhaps noticed the following
apparent discrepancy here: while Sorkin's $P_{i}$s were purely
discrete {\em topological} structures, their Gel'fand-dual picture
in terms of the $\omg_{i}$ appears to encode additional
information about the {\em differential} structure (of the
original continuum $X$ that Sorkin started with). How did
`differentiability' (differential structure) creep into our
considerations when, following Sorkin, the original investigations
pertained only to `continuity' (:topological structure)? The
reason is that the $P_{i}$s can be also thought of as {\em
homological} objects---as a matter of fact, as {\em simplicial
decompositions} of the original manifold $X$. That is to say, the
$P_{i}$s can alternatively (and equivalently) be viewed as {\em
simplicial complexes} $\simp_{i}$, and as a result, their
corresponding incidence algebras as {\em incidence algebras of
simplicial complexes} $\omg_{i}(\simp_{i})$
\cite{rapzap1,rapzap2,zap1}.\footnote{Indeed, the order $n$ of
each $\omg^{n}$ in (\ref{eq17}) corresponds to the {\em simplicial
degree} (or cardinality) of the respective simplex in $\simp_{i}$
\cite{rapzap1}.} The $d_{i}$s of the $\omg_{i}$s can now be
expressed in terms of the nilpotent homological boundary $\delta$
(border) and coboundary $\delta^{*}$ (coborder) operators
\cite{zap1}.\footnote{In categorical terms, the simplicial
analogue of the correspondence (\ref{eq16}),
$\simp\mapto\omg_{i}$, turns out to be a (contravariant) functor
between the category of (finitary) simplicial complexes and
simplicial maps (or equivalently, the category of finitary posets
and poset morphisms---{\it ie}, order preserving/monotone maps),
and the category of (finitary) incidence algebras and algebra
homomorphisms \cite{rapzap1,rapzap2,zap1}.} The dual character of
the $\omg_{i}$s relative to the $\simp_{i}$s can now be understood
simply by noting that the former's elements are {\em
cohomological} entities---{\it ie}, discrete differential
form-like objects, which are obviously dual to the homological
simplices in the $\simp_{i}$s.

\paragraph{Finitary differential triads.} The observation above
that the $\omg_{i}$s encode not only topological, but also {\em
differential geometric} information (coming from $X$), motivated
this author to try to apply the ADG-machinery to a finitary
setting. But for that, some {\em sheaf}-like structure was needed
to be introduced first.\footnote{The motivation mentioned above
was a {\em mathematical} one. The {\em physical} motivation was
that this author ultimately wished to localize or gauge (thus
dynamically vary and `curve') quantum causality ({\it ie}, the
incidence algebras modelling qausets) \cite{malrap1,malrap3,rap5}.
In turn, the act of `localization' or `gauging' is
(mathematically) tautosemous to `sheafification' \cite{mall8},
followed by endowing the resulting sheaf with a connection $\conn$
\cite{mall1,mall2}.} Thus, {\em finitary spacetime sheaves}
(finsheaves) $\mathcal{S}_{i}$ over Sorkin's finitary posets were
introduced and developed in \cite{rap2}. Originally, finsheaves
were conceived, in complete analogy to the $P_{i}$s, as genuine
finitary approximations of the sheaf $\cont_{X}$ of continuous
($\R$-valued) functions on the topological manifold $X$, again in
the sense that an inverse system thereof possessed a projective
limit sheaf that is topologically indistinguishable from
$\cont_{X}$. However, the original intention to build {\em
differential}, not just topological, structure into the finsheaves
mandated that this author should define {\em finsheaves of
incidence algebras}. This definition was straightforward to arrive
at since it was realized early on that the map (\ref{eq16}) is, by
construction,\footnote{The construction alluded to above was
coined {\em Gel'fand spatialization} in \cite{rapzap1,rapzap2}
(see also \cite{zap0}), whereby roughly, the `local' Sorkin
order-topology of $P_{i}$ is equivalent to the `local' Rota
topology assigned to the (primitive) spectra of the $\omg_{i}$s, a
procedure which is effectively an application of Gel'fand duality
to the finitary realm of the $P_{i}$s.} a {\em local
homeomorphism}---{\it alias}, a {\em sheaf} \cite{mall1,mall2}.
Thus finsheaves of incidence algebras $\Omg_{i}$---essentially,
the sheaf-theoretic localizations of the $\omg_{i}$s over Sorkin's
$P_{i}$s---were introduced, and hence the ADG-theoretic panoply
was ready to be used in the finitary realm.

Indeed, finsheaves of incidence algebras define (graded) {\em
finitary differential triads}

\begin{equation}\label{eq18}
\triad_{i}:=(\struc_{i}\equiv
\aconn_{P_{i}},d_{i},\mathbf{R}_{i}\equiv\mathcal{R}_{P_{i}}=\bigoplus_{n\geq1}\Omg^{n}_{i})\footnote{Again,
like before, from now on we will omit the base space $P_{i}$
subscript from the finsheaves involved, but we will retain the
`finitarity' or resolution index `$i$' to be used in the
projective and inductive limits subsequently. Also note that built
into $\triad_{i}$ are higher order (or grade) extensions
$\Omg^{n}$ of the $\Omg^{1}$ appearing in the abstract
differential triad in (\ref{eq4}), as well as higher order
prolongations $d_{i}^{n}~(n\geq1)$ of $\partial_{i}\equiv
d_{i}^{0}$, which $\mathbf{K}$-linearly map $\Omg_{i}^{n}$ to
$\Omg_{i}^{n+1}$ \cite{malrap1,malrap2,malrap3}. The latter will
be generically represented by the finitary version $d_{i}$ of the
Cartan-de Rham-K\"ahler (exterior) differential.}
\end{equation}

\noindent which have been seen to carry, virtually unaltered, to
the `discrete', finitary setting certain key results of the CDG of
smooth manifolds, such as the {\em Poincar\'e lemma}, the {\em de
Rham theorem}, the {\em Weil integrality theorem}, the {\em
Chern-Weil theorem}, and much more (pertaining, for example, to
geometric prequantization of gravity in an ADG-setting)
\cite{malrap2}.

Those applications aside for a moment, at this point we would like
to close this paragraph by giving a characteristic example of the
aforementioned categorical versatility of ADG, as opposed to the
rigidity of the manifold based CDG and of the category $\man$
underlying it. To this end, we show how one can arrive
straightforwardly from Sorkin's finitary posets to finitary
differential triads without having to go the long laboriously
`constructive' way via simplicial complexes, their Gel'fand-dual
incidence algebras and the finsheaves thereof.\footnote{The reader
should note that in the past trilogy
\cite{malrap1,malrap2,malrap3} of finitary applications of ADG, we
indeed followed that `roundabout' way in order to define finitary
differential triads.} This involves an immediate application to
the Sorkin scheme of the push-out and pull-back (along continuous
maps between base topological spaces) results mentioned in the
previous section, as follows:

\begin{itemize}

\item First, unlike Sorkin whose considerations in \cite{sork0} were purely topological as noted earlier,
we assume that (the region of) the manifold $X$ carries not only
the usual topological ($\cont$), but also the standard
differential ($\smooth$) structure of a locally Euclidean space.
What amounts to the same from an ADG-theoretic vantage, we suppose
that $X$ supports the classical $\mathbf{K}$-algebraized space
$\diad_{\infty}:=(X,\smooth_{X})$ carrying the classical
differential triad $\triad_{\infty}:=(\smooth_{X},\partial
,\Omg_{X})$ of a differential manifold.

\item Then, we factor {\it \`a la} Sorkin $X$ by
$\stackrel{\gauge_{i}}{\sim}$ to obtain the finitary substitute
$P_{i}$ (\ref{eq13}) and, as a result, the continuous surjection
$f_{i}$ between them (\ref{eq15}).

\item Finally, we evoke the push-out result of Papatriantafillou
and endow $P_{i}$ with the differential triad
$f_{i*}(\triad_{\infty})$, which can be readily identified with
the finitary differential triad $\triad_{i}$ of (\ref{eq18}).

\end{itemize}

\noindent This $f_{i*}$-induction of the (essentially
algebraic---{\it ie}, sheaf-theoretic) differential geometric
mechanism from $\triad_{\infty}$ on the continuum $X$ to
$\triad_{i}$ on the `discretum' $P_{i}$ has been recently coined
the `{\em Newtonian spark}' in \cite{mall10,malrap4} and it
exemplifies what we regard as being {\em the} subtle epitome of
ADG, namely, that although we may initially inherit the
(essentially algebraic) differential geometric mechanism---in
essence, the differential $d$---from a base space (here, be it a
locally Euclidean one such as $X$), we then `forget' about that
background sheaf-theoretic `localization scaffolding' and develop
all the various differential geometric constructions
`algebraically in the stalk' ({\it ie}, with the algebraic objects
living in the relevant sheaves' spaces---or what is the same,
solely with the relevant sheaves' sections), and what's more,
completely independently of that surrogate $X$, which just
furnished us with the invaluable for actually doing {\em
differential} geometry $d$.

\paragraph{Finitary vacuum Einstein equations.} It has been shown
\cite{malrap3} that with the $\triad_{i}$s and the general
ADG-machinery in hand, one can transcribe to the finitary realm
all the ideas and constructions of the manifold based
(pseudo-)Riemannian geometry that we recalled in section 2. That
is, one can develop a `{\em finitary Riemannian geometry}', which
is a particular instance of the background manifoldless Riemannian
geometry of section 2. In particular, one can formulate on each
$\triad_{i}$ a finitary version of the vacuum Einstein equations
(\ref{eq9}), reading:

\begin{equation}\label{eq19}
\ricci_{i}(\modl_{i})=0
\end{equation}

\noindent with $\ricci_{i}$ the finitary version of the Ricci
scalar, and $\modl_{i}$ the $*$-dual of the finsheaf $\Omg_{i}$ of
incidence algebras, as posited by ADG.

Having the $\triad_{i}$s and (\ref{eq19}) holding on each of them
at our disposal, in the next section we take on their inverse and
direct limits.

\section{The Category of Differential Triads is Bicomplete}

That $\triad$ is bicomplete is in fact just a result (theorem) in
the category-theoretic perspective on ADG
\cite{pap1,pap2,pap3,pap4,pap5}, but due to its importance in the
present paper, we promote it to the title of the present section.
Indeed, as noted earlier, $\triad$ is closed under both projective
and inductive limits. This means that inverse and direct systems
of differential triads possess categorical limit and colimit
spaces that are themselves differential triads. Since
$\ctriad_{i}$ is a subcategory of $\ctriad$, projective and
inductive systems of $\triad_{i}$s have inverse and direct limit
structures that are themselves triads; albeit, not necessarily
{\em finitary}, `discretum' ones. In fact, as we shall see in the
next paragraph, the limit triads that we are interested in and are
of significance for the physical interpretation of our theoretical
scheme are `{\em infinitary}', `{\em classical continuum}' ones.

\paragraph{Inverse and direct limits of $\triad_{i}$s and their vacuum
Einstein equations.} We first start with Sorkin's result noted
earlier, namely, that the inverse system $\inv$ of the $P_{i}$s
has, at the limit of maximum (topological) resolution or
refinement of the $\gauge_{i}$s, a projective limit space
$P_{\infty}$ that for all intents and purposes is topologically
indistinguishable from ({\it ie}, homeomorphic to) the
$\cont$-continuum $X$ that we started with. Likewise for the
analogous finsheaf-discretizations of $\cont_{X}$ in \cite{rap2}.

As it has already been pointed out numerous times in the past
trilogy \cite{malrap1,malrap2,malrap3}, since the $\omg_{i}$s are
categorically dual to the $P_{i}$s, one infers that they too
comprise, dually now, a {\em direct} system $\diromg=\{\omg_{i}\}$
possessing an inductive limit incidence algebra which, in view of
the fact that the $\omg_{i}$ encode information not only about the
topological, but also about the differential, structure of the
continuum $X$, should come close to emulating the classical
differential geometric structure of $X$---namely, the
$\smooth(X)$-module of differential forms on the differential
manifold $X$.\footnote{This has been investigated in detail in
\cite{rapzap1,rapzap2,zap1,zap2}.} Accordingly, passing to
finitary (`discretum') differential triads, they also constitute a
projective/inductive system $\invtriad$\footnote{The joint epithet
`projective/inductive' to $\invtriad$ pertains exactly to the
duality mentioned above: while the $P_{i}$s---the base spaces of
the $\triad_{i}$s---constitute an inverse system $\inv$, their
categorically dual $\omg_{i}$s---inhabiting the stalks of the
finsheaf spaces in the $\triad_{i}$s---constitute a direct system
$\diromg$. Informally-syntactically speaking,
$\gauge_{i}$-refinement for the $P_{i}$s goes from-right-to-left,
while for the $\omg_{i}$s from-left-to-right. (See expression
(150) in \cite{malrap3}.)} possessing, according to
Papatriantafillou's results above, at the infinite limit of
resolution (refinement) of the $\gauge_{i}$s, an `infinitary'
(continuum) triad $\striad$ which comes as close as possible (via
Sorkin's scheme) to the classical one
$\triad_{\infty}=(\smooth_{X},\partial ,\Omg_{X})$ supported by
the differential manifold $X$.

The expression `comes as close as possible to $\triad_{\infty}$'
above pertains to the fact that, much in the same way that one
does not actually recover $X$ as the inverse limit space of
$\inv$, one also does not exactly get $\smooth_{X}$ and the
$\smooth(X)$-module sheaf $\Omg$ of (germs of) smooth differential
forms (over the differential manifold $X$'s points) at the direct
limit of (infinite localization of) the $\omg_{i}$s in $\diromg$.
Rather, similarly to the fact that one gets a `larger' inverse
limit topological space $P_{\infty}$ having $X$ as a dense subset
in Sorkin's scheme ({\it ie}, roughly, $P_{\infty}$ has more
points than the original $X$), one anticipates $\diromg$ to yield
at the inductive limit a (`topological') algebra $\aconn_{\infty}$
`larger' than $\smooth(X)$ and consequently an
$\aconn_{\infty}$-module $\mathcal{R}_{\infty}$ of differential
form-like entities `larger' than the standard $\smooth$-one. In
Zapatrin's words, when he was working out continuum limits of
incidence algebras of simplicial complexes \cite{zap1,zap2}:
``{\em it is as if too many functions and forms want to be smooth
in the continuum limit}''.\footnote{Roman Zapatrin in private
e-mail correspondence.} One intuits that much in the same way that
Hausdorff reflection gets rid of the `extra points' of
$P_{\infty}$ to recover the $\cont$-manifold $X$, so by ridding
$\omg_{\infty}$ of the `rogue' extra functions and forms on
$P_{\infty}$ ({\it eg}, by factoring it by a suitable differential
ideal \cite{zap1,zap2}), one should recover the usual smooth
functions and forms over the differential manifold $X$.
Nevertheless, the important point for the exposition here is that
{\em one does indeed get a continuum differential triad}, which
however, only in order to be formally distinguished from the
classical $\smooth$-smooth one $\triad_{\infty}$ to avoid any
minor technical misunderstanding, one might wish to call
`$\ssmooth$-\texttt{smooth}' and symbolize it by $\striad$
\cite{malrap3}. On the other hand, after having alerted the reader
to this slight distinction between $\striad$ and
$\triad_{\infty}$, in the sequel, for all practical purposes and
in order to avoid proliferation of redundant symbols, we shall
abuse language and assume that $\striad$ and $\triad_{\infty}$ are
`essentially isomorphic' ({\it ie}, effectively equivalent and
indistinguishable). So, both will be generically referred to as
{\em the classical continuum differential triad} (CCDT), with the
symbols $\triad_{\infty}$ and $\striad$ used interchangeably.

Thus, we formally write for this joint inverse/direct limit
procedure exercised on $\invtriad$

\begin{equation}\label{eq20}
\underset{\infty\leftarrow
i}{\overset{i\rightarrow\infty}{\lim}}\invtriad=\striad\equiv\triad_{\infty}
\end{equation}

\noindent As argued and shown in detail in \cite{malrap3}, each
$fcq$-differential triad $\triad_{i}$ carries on its shoulders the
whole ADG-machinery and structural panoply involved in the usual
manifold based (pseudo)-Riemannian geometry. In particular,
(\ref{eq19}) holds on each $\triad_{i}$ and hence this information
carries to the inverse/direct limit of $\invtriad$, which in turn
is seen to support an inverse system $\inveinst$ of $fcq$-vacuum
Einstein equations.\footnote{For example, again see expression
(150) in \cite{malrap3}.} We thus recover a smooth continuum limit
version of the vacuum Einstein equations, holding on $\striad$ (or
equivalently, on $\triad_{\infty}$), which we formally depict as:

\begin{equation}\label{eq21}
\underset{\infty\leftarrow
i}{\lim}\inveinst=\underset{\infty\leftarrow
i}{\lim}\ricci_{i}(\modl_{i})=\ricci_{\infty}(\modl_{\infty})=0
\end{equation}

\noindent with $\modl_{\infty}$ the dual of the
$\smooth_{X}$-module sheaf of (germs of) smooth differential forms
on the differential manifold $X$ comprising the CCDT
$\triad_{\infty}$, and $\ricci_{\infty}$ the classical smooth
({\it ie}, $\struc\equiv\smooth_{X}$-valued) Ricci curvature
scalar.

\paragraph{`Correspondence limit/principle' interpretation of
inverse/direct limits.} We briefly remark here that in
\cite{rapzap1,rapzap2}, in view of the quantum interpretation that
the $\omg_{i}$s enjoy, the continuum inverse limit of $\inv$, and
dually, the direct limit of $\diromg$, was interpreted as {\em
Bohr's correspondence principle}, otherwise known as {\em the
classical continuum limit}. As a result, (\ref{eq21}) may be
interpreted as follows: {\em at the continuum limit of infinite
topological resolution (or refinement) of $X$ into its points, or
equivalently, of infinite sheaf-theoretic localization of the
incidence algebras over $X$'s points, one obtains the classical
continuum vacuum Einstein equations\footnote{That is, the vacuum
Einstein equations holding over the entire smooth manifold
$X$---{\it ie}, on $\triad_{\infty}$.} from the individual
$fcq$-ones} {\em holding on each `discretum' triad $\triad_{i}$}
(\ref{eq19}). In other words, and this will prove to be of
importance for some remarks that we are going to make in the next
two sections regarding the application of ADG to both classical
and quantum gravity, {\em ADG, and the vacuum Einstein gravity to
which it has been applied so far, is genuinely background
spacetime independent, {\it ie}, the vacuum Einstein equations are
in force independently of whether one assumes the base space(time)
to be a `quantal discretum'\footnote{As it were, when (locally at
least) only a finite number of `degrees of freedom' of the vacuum
gravitational field are excited ({\it ie}, when only a locally
finite number of events are involved, or dually/functionally, when
only a finite number of `modes' of the gravitational field
`contribute' to/`participate' at the gravitational dynamics at
each spacetime event), and when some sort of quantization has
already taken place \cite{rapzap1,rapzap2}.} or a `classical
continuum'}.\footnote{That is, when the gravitational field
`triggers' or `excites' a continuous infinity of spacetime events
in the manifold, and all `quantum interference' (coherent quantum
superpositions between the elements of the $\omg_{i}$s) has been
lifted \cite{rapzap1,rapzap2}.}

\section{Finitary-Algebraic Evasion of the Interior Schwarzschild Singularity}

We have now built a sufficient conceptual and technical background
to present in a straightforward fashion the ADG-theoretic evasion
of the inner Schwarzschild singularity entirely by finitistic and
algebraic means. First however, in order to present that
`resolution' in a more effective way, we recall a contrasting
theory of the interior Schwarzschild singularity. This is the
standard one based on the usual approach to GR via CDG, the
$\smooth$-smooth base spacetime manifold and the smooth Lorentzian
metric on it ({\it ie}, {\it in toto}, the classical
pseudo-Riemannian geometry underlying GR). The following are well
known, amply worked out facts about the Schwarzschild solution of
the Einstein equations, which we thus present rather briefly and
informally.

\paragraph{Classical Schwarzschild preliminaries: the standard view, the usual suspects and the main
problematics.} We begin by noting some familiar features of GR.
First of all, the original theory was formulated in terms of the
smooth metric tensor $g_{\mu\nu}$ on a differential spacetime
manifold $X$. That is to say, the sole dynamical variable in GR
(as originally formulated by Einstein) is $g_{\mu\nu}$, whose ten
independent $\smooth(X)$-valued components represent the {\em
gravitational potentials} and at the same time they enter into the
pseudo-Riemannian line element representing the {\em
chrono-geometric structure} of the spacetime continuum. In a
nutshell, $g_{\mu\nu}$ represents gravity-{\it cum}-background
spacetime chronogeometry, and the dynamical equations that it
obeys in the absence of matter are the (vacuum) Einstein equations
(\ref{eq21})---formally, non-linear (hyperbolic) second-order
partial differential equations (PDEs) for $g_{\mu\nu}$.

The Schwarzschild solution of the said equations represents the
spherically symmetric vacuum gravitational field outside a
massive, spherically symmetric body of mass $m$. On grounds of
physical import alone, our choice of this particular solution on
which to exercise our ADG-machinery and results in order to
`resolve' it may be justified on the fact that experimentally all
the differences between non-relativistic (Newtonian) gravity and
GR have been based on predictions by this solution \cite{hawk0}.
Also, since comparison with Newtonian gravity allows us to
interpret the Schwarzschild solution as the gravitational field
(in empty spacetime) produced by a point-mass source $m$ viewed
from far away ({\it ie}, from infinity) \cite{hawk0},
Finkelstein's original treatment of the Schwarzschild
gravitational field as being produced by a point-particle in an
otherwise empty spacetime manifold \cite{df} appears to be a good
starting choice.

So first, following Finkelstein, one assumes that spacetime is a
smooth ($\smooth$) or even an analytic ($\anal$) manifold
$X$,\footnote{In this paper we shall {\em not} distinguish between
a $\smooth$- and a $\anal$-manifold (or for the same reason,
between CDG and Calculus or Analysis). From an ADG-theoretic
viewpoint, as noted earlier, a smooth manifold $X$ corresponds to
choosing $\smooth_{X}$ for structure sheaf, while an analytic one
has $\struc\equiv\anal_{X}$---the structure sheaf of coordinate
functions (of $X$'s points) each admitting analysis (expansion)
into power series. Admittedly, $\anal$- is a slightly stronger
assumption for a manifold than $\smooth$-, but this does not
change or inhibit the points we wish to make here about the
Schwarzschild singularity and its bypass in the light of ADG.} and
then one places at its `center' (:interior) a point-mass $m$. For
a Cartesian coordinate system with $m$ at its origin, the
`effective' spacetime manifold of this point-particle becomes $X$
minus the particle's `wristwatch' time-line $L_{t}:=\{ p\in
X:~x_{i}(p)=0,~(i=1,2,3,~t\equiv x_{0})\}$ ; that is to say,

\begin{equation}\label{eq22}
X_{S}=X-L_{t}
\end{equation}

\noindent with the subscript `$S$' standing for
`($S$)chwarzschild'. Then, one observes that $m$ is the source of
a gravitational field, represented by a smooth (or analytic)
spacetime metric $g_{\mu\nu}$, which obeys the vacuum Einstein
equations (\ref{eq21}). The Schwarzschild solution of the
equations (\ref{eq21}) is the Schwarzschild metric
$g_{\mu\nu}^{S}$ expressed in Cartesian-Schwarzschild coordinates,
which in turn defines an infinitesimal proper time increment, as
follows:

\begin{equation}\label{eq23}
ds_{S}^{2}=(1-r_{S}^{-1})(dx_{S}^{0})^{2}-(1-r_{S}^{-1})^{-1}dr_{S}^{2}-(dx_{S}^{i}dx_{S}^{i}-dr_{S}^{2})
\end{equation}

\noindent expressed in normalized, `natural' units in which the
so-called Schwarzschild radius ($r=2m$) and the speed of light
($c=10^{8}m/s$) are equal to $1$.\footnote{Also, in (\ref{eq23})
above, $r_{S}=\sqrt{x_{S}^{i}x_{S}^{i}}$ and
$dr_{S}=r_{s}^{-1}x_{S}^{i}dx_{S}^{i}$. The more familiar ({\it
ie}, not in natural units) expression for the Schwarzschild line
element in cartesian coordinates is
$(1-\frac{2m}{r})dt^{2}+dx^{2}+dy^{2}+dz^{2}+\frac{2m}{r(r-2m)}(xdx+ydy+zdz)^{2}$,
while in spherical-Schwarzschild coordinates (again not in natural
units), it reads
$-(1-\frac{2m}{r})dt^{2}+(1-\frac{2m}{r})^{-1}dr^{2}+r^{2}(d\theta^{2}+\sin^{2}\theta
d\phi^{2})$.}

Evidently, $g_{\mu\nu}^{S}$ has two singularities: one right at
the {\it locus} of the point-mass---the Cartesian origin ($r=0$),
and one at the Schwarzschild radius ($r=1$) delimiting a spacelike
$3$-dimensional unit-spherical shell in $X$, commonly known as the
Schwarzschild horizon. The two singularities are usually pitched
as the {\em interior} (inner) and {\em exterior} (outer)
Schwarzschild singularities, respectively.\footnote{The
Schwarzschild horizon is the horizon of the Schwarzschild black
hole, and it is supposed to host the inner Schwarzschild
singularity at its kernel, `beyond the horizon'.}

In \cite{df}, Finkelstein initially considered an analytic metric
$g_{\mu\nu}^{F}$ on $X$, expressed in what is nowadays usually
called Eddington-Finkelstein coordinates,\footnote{The
Eddington-Finkelstein frame consists of so-called {\em
logarithmic-null spherical coordinates} $(n^{\pm},r,\theta
,\phi)$, with the null coordinate $n^{\pm}$ being either {\em
advanced} $n^{+}:=t+r^{'}$ or {\em retarded} $n^{-}:=t-r^{'}$, and
$r^{'}$ defining a logarithmic radial coordinate
$r^{'}:=\int\frac{dr}{1-2mr^{-1}}=r+2m\log(r-2m)$.} defining the
following infinitesimal spacetime line element

\begin{equation}\label{eq24}
\begin{array}{c}
ds^{2}_{F}=(1-r_{F}^{-1})(dx_{F}^{0})^{2}+2r_{F}^{-1}dx_{F}^{0}dr_{F}-
(1+r_{F}^{-1})dr_{F}^{2}-(dx_{F}^{i}dx_{F}^{i}-dr_{F}^{2})=\cr
-(1-\frac{2m}{r})(dn^{\pm})^{2}\pm
2dn^{\pm}dr+r^{2}(d\theta^{2}+\sin^{2}\theta d\phi^{2})
\end{array}
\end{equation}

\noindent and he then showed that, for the region of $X$ outside
the Schwarzschild horizon $3$-shell ($r_{F}>1$), the following
simple `logarithmic time coordinate change' from the analytic
Finkelstein ${}^{\omega}\struc_{F}={}^{\omega}(x^{\mu}_{F})$
coordinates to the also analytic Schwarzschild ones
${}^{\omega}\struc_{S}={}^{\omega}(x^{\mu}_{S})$

\begin{equation}\label{eq25}
\begin{array}{c}
\underline{{}^{\omega}\struc_{F}\mapto{}^{\omega}\struc_{S}:}\cr
\cr x_{F}^{0}\mapto x_{S}^{0}=x_{F}^{0}+\mathrm{ln}(r_{F}-1)\cr
x_{F}^{i}\mapto x_{S}^{i}=x_{F}^{i}
\end{array}
\end{equation}

\noindent transforms the line element $ds^{2}_{F}$ (and the
associated $g_{\mu\nu}^{F}$) in (\ref{eq24}) to the Schwarzschild
one $ds_{S}^{2}$ (and its associated $g_{\mu\nu}^{S}$) in
(\ref{eq23}).

Conversely, he argued that since $\ricci_{\infty}$ in (\ref{eq21})
is a tensor with respect to the $^{\omega}\struc_{F}$ coordinates,
the vacuum Einstein equations hold in all $X$ (now coordinatized
by ${}^{\omega}\struc_{F}$\footnote{Which we may just as well
symbolize by $X_{F}$.})---in particular, they hold on the
Schwarzschild horizon unit-shell.

{\it In toto}, Finkelstein showed that the analytic coordinate
change

\begin{equation}\label{eq26}
X_{S}\equiv(X,{}^{\omega}\struc_{S})\mapto(X,{}^{\omega}\struc_{F})\equiv
X_{F}
\end{equation}

\noindent amounts to an {\em analytic extension} of $X_{S}$
(coordinatized by the Cartesian ${}^{\omega}\struc_{S}$ and
carrying the analytic $g_{\mu\nu}^{S}$ defining $ds_{S}^{2}$
above---which is singular at $r=1$), to $X_{F}$ (coordinatized by
the analytic ${}^{\omega}\struc_{F}$ and carrying the analytic
$g_{\mu\nu}^{F}$ defining $ds^{2}_{F}$, which is not singular at
the Schwarzschild radius!).

In fact, Finkelstein showed that the said analytic extension of
$X_{S}$ to $X_{F}$ can be carried out in two distinct
ways,\footnote{Depending on whether one chooses advanced or
retarded logarithmic-null coordinates.} each one being the
time-reversed picture of the other, which in turn means that the
$r=1$ Schwarzschild horizon, far from being a real singularity,
acts as ``{\em a true unidirectional membrane}'' in the sense that
``{\em causal influences can pass through it only in one
direction}'' and, moreover, he gave a particle-antiparticle
interpretation of this gravitational time-asymmetry
\cite{df}.\footnote{The null (in the Finkelstein frame)
hypersurface Schwarzschild horizon is also known as a {\em closed
trapped surface} \cite{hawk0}, which `traps' past- (resp. future-)
directed causal ({\it ie}, timelike or null) signals depending on
whether one chooses advanced (resp. retarded) Finkelstein
coordinates to chart the original manifold. Also, it can be easily
seen that inside Schwarzschild horizon the original time and
radial coordinates exchange roles.}

On the other hand, about the inner Schwarzschild singularity
Finkelstein concluded that the theory ({\it ie}, the manifold and
CDG-based GR) is out of its depth as there is no (analytic)
coordinate change that can remove it like the outer one. In other
words, the interior Schwarzschild singularity, right at the
point-particle $m$, is regarded as being a `genuine', `true'
singularity of the gravitational field, not removable (or
resolvable) by analytic ({\it ie}, CDG-theoretic) means
\cite{df,hawk0,clarke4}.\footnote{Indeed, in the $n^{+}$-picture,
any future-directed causal curve crossing the Schwarzschild
horizon can reach $r=0$ in finite affine parameter distance (see
next paragraph). Moreover, it can be shown that as $r\mapto 0$ the
Ricci scalar curvature $\ricci$ in (\ref{eq21}) blows up as
$\frac{m^{2}}{r^{6}}$, while there is no further analytic
extension (even in a $C^{2}$-differential, or even in a
$C^{0}$-topological, fashion!) of the Schwarzschild spacetime
manifold across the $r=0$ {\it locus}.} Which brings us to the
general consensus about `true', as opposed to merely `virtual' or
`coordinate', $\smooth$-spacetime singularities.

\paragraph{`True' versus `coordinate' singularities: a CDG-conservatism and monopoly
underlying all approaches (so far) to spacetime singularities.}
The two Schwarzschild singularities above provide a good example
of the general way we think about and actually deal with
gravitational singularities in the CDG and manifold based GR.

To begin with, it must be stressed up-front that {\em there is no
general, concise and `rigorous' definition of (`true')
singularities in GR} \cite{geroch,hawk0,clarke4,rendall}. Rather,
one proceeds {\em by elimination and exclusion} in order to
identify genuine gravitational singularities and separate them
from `apparent', coordinate ones, in the following way. Given a
singular gravitational spacetime---by which one means a manifold
$M$ (of a certain order of differentiability\footnote{That is, an
analytic ($\struc\equiv\anal_{M}$), or smooth ($\smooth_{M}$), or
even a manifold of finite order of differentiability
($C^{k}_{M}$).}) endowed with a (Lorentzian) metric $g$ (of
maximum order of differentiability assumed for the underlying $M$)
satisfying Einstein's equations and possessing singularities at
certain {\it loci} of $M$---one tries to analytically (or anyway,
smoothly, or in a $C^{k}$-fashion) extend $M$\footnote{Here, `to
extend $M$' means essentially `{\em to change coordinate structure
sheaf of differentiable functions on $M$}'.} past those {\it loci}
so as to include them with the other `regular' points of $M$. If
there happens to be such an extension, the singularity in focus is
regarded as an `{\em apparent}', `{\em virtual}', {\em coordinate}
one---an indication that the physicist originally chose an
inappropriate system of coordinates (patches) to chart $M$ and to
express $g_{\mu\nu}$ with respect to it. The exterior
Schwarzschild singularity we saw earlier is the archetypical
example of such a coordinate singularity. On the other hand, if
there is no such extension, the `anomalous' {\it locus} is branded
a `{\em true}', `{\em real}', `{\em genuine}' singularity. The
inner Schwarzschild singularity is the archetypical example of
such a real singularity, in the vicinity of which $g_{S}$ (and the
Ricci scalar) diverges to infinity. Kruskal's maximal analytic
extension of $X_{S}$ above did not manage to include it with the
other regular points of the spacetime manifold \cite{kruskal}.
Coordinate singularities are {\em not} considered to be `{\em
physical singularities}' ({\it ie}, they are not of physical
significance), while genuine ones {\em are}
\cite{geroch,hawk0,clarke4,rendall}.

Clearly then, coordinate singularities are regarded as being `{\em
regular points in disguise}', and the differential manifold,
together with the differential equations of Einstein that it
supports, are still in force since they can be continued past
them. On the other hand, this is not so for true singularities.
The latter are {\it loci} where the differential law of gravity
appears to stop ({\it ie}, it ceases to hold) somehow, or even
more graphically, it {\em breaks down}. Genuine singularities are
sites where CDG (and the smooth manifold supporting its
constructions) has reached the limit of its applicability and
validity. Thus, let us recall briefly from \cite{hawk0,clarke4}
the three general kinds of gravitational singularities, and what
underlies them all. We shall first mention {\it en passant} how
one usually copes with genuine singularities in a manifestly
CDG-conservative fashion.

Apart from {\em analytic inextensibility}, the other `defining'
feature of real spacetime singularities is {\em (causal geodesic)
incompleteness}. Roughly, the idea behind {\em spacetime
incompleteness} is that (material) particles cannot reach
(genuine) singularities in `finite (proper) time' by following,
under the focusing action of the strong gravitational field at the
purported singular {\it loci}, smooth (causal) paths (geodesics)
in the manifold $M$. Historically, the importance of
(causal---{\it ie}, timelike and null) geodesic incompleteness was
first recognized in \cite{geroch}. Subsequently, null geodesic
incompleteness was the central prediction of the celebrated
singularity theorems of Hawking and Penrose \cite{hawk1,hawk0}.
However, it is not entirely clear what spacelike incompleteness
means physically, since spacelike curves in $M$ do not have an
interpretation as histories of physical objects ({\it ie}, fields
and their particles). On the other hand, as Clarke points out in
\cite{clarke4}, one need not consider only `free falling'
observers following causal {\em geodesics}, since other physically
admissible frames---ones with bounded acceleration for
instance---may be able to reach the point-{\it loci} in question
in finite (proper) time, even though geodesic observers cannot. In
order to include the world-lines of such in principle arbitrarily
accelerated observers, paths more general than geodesics---ones
parameterized not by proper time, but by an arbitrary so-called
{\it general affine parameter}---must also be included in the
definition of incompleteness. {\it In toto}, incompleteness
pertains to the idea that curves of finite (general affine
parameter) length cannot reach the singular {\it loci} in
question. The bottom-line of all this is that {\em a spacetime is
called singular if it is incomplete and inextensible}, in the
above sense.

Now there appears to be a clear-cut way to proceed in dealing with
true spacetime singularities, namely, one can relegate them to the
`edge' of a maximally extended spacetime manifold and view them as
`asymptotically terminal points' of incomplete curves. That is,
one thinks of genuine singularities as {\it loci} situated on a
certain boundary set $\partial M$ adjoined to $M$, with the latter
endowed with an `appropriate' topology, which in turn qualifies
$\overline{M}=M\cup\partial M$ as {\em the closure of $M$} and
recognizes $\partial M$ as a topological boundary proper.
Parenthetically, without going into any detail, so far there are
two basic singular boundary constructions: the {\em causal
boundary} of Geroch, Kronheimer and Penrose \cite{geroch2}, and
the so-called {\em b-boundary} of Schmidt \cite{schmidt}. Each of
these two boundaries (and associated topologies) has its own pros
and cons that we do not want to go into here, but for a detailed
exposition of and comparison between them the reader is referred
to \cite{hawk0,clarke4}.

Having ascribed a topology and a boundary to the spacetime
continuum, and concomitantly having pushed the genuinely singular
{\it loci} out of the regular $M$ and virtually to `the margin of
spacetime' ({\it ie}, onto $\partial M$), one identifies three
general types of genuine $\smooth$-gravitational singularities
\cite{clarke3}:

\begin{enumerate}

\item \ovalbox{\bf (Differential) geometric singularities (DGS):} boundary points
for which there is no $C^{k}$-differential extension of (the
metric on) $M$ so as to remove them.

\item \ovalbox{\bf (Various) energy singularities (VES):} boundary
points for which there is no (analytic) extension of $M$ that
removes them satisfying at the same time various energy conditions
(inequalities) \cite{clarke4}, the most prominent and generic ones
being {\em gravitational energy positivity} (:gravity is always
attractive) and the associated {\em weak and dominant energy
conditions} \cite{hawk0}.

\item \ovalbox{\bf (Solution) field singularities (SFS):} boundary points for which
there is no (analytic) extension of $M$ that removes them and is a
solution of the Einstein field equations in question ({\it eg},
Einstein-scalar or the Einstein-fluid equations). The important
thing to mention here is that the term {\em solution} to the field
equations means {\em generalized smooth} or {\em smeared}---what
is commonly known as {\em distributional}, solution.

\end{enumerate}

\noindent Plainly, {\em (analytic) inextensibility}---loosely
speaking, our inability to apply CDG or Analysis---{\em underlies
all three `definitions' of genuine singularities above}.
Metaphorically speaking, {\em true singularities are breakdown
points of the Differential Calculus}. We will come back to this
point in the sequel.

In the present paper we will be predominantly interested in
DGSs---which incidentally are singularities of the `purest' kind
{\it vis\`a-vis} differential geometric considerations
\cite{clarke3}---as they manifestly depict the aforesaid Calculus
or `classical differentiability' breakdown, as Clarke explicitly
points out in \cite{clarke3}:

\begin{quotation}
\noindent ``{\small ...Thus the definition of a {\rm [differential
geometric]} singularity depends on the definition of an {\rm
[analytic]} extension of {\rm [the]} space-time {\rm [manifold]},
and so the question of what counts as a singularity depends on
what sort of extension is allowed. We call a boundary point {\rm
[of a smooth manifold]} a class $C^{k}$ {\rm [differential]}
geometrical singularity if there is no {\rm [analytic]} extension
with a $C^{k}$ metric that removes it; i.e. if it is associated
with a breakdown of differentiability of the metric at the $C^{k}$
level...''}\footnote{In square brackets are our own additions for
clarity and completeness.}
\end{quotation}

It must be also noted here that the way in which we ascribe a
topology and construct a boundary to $M$ on which true
singularities are located, apart from its physical
motivation,\footnote{For example, understandably the physicist
would like to have a `controlled' study of the asymptotic behavior
of, say, the Riemann curvature tensor (whose components represent
gravitational tidal forces) as one approaches ({\it ie}, in the
immediate neighborhood of) the singularity in focus---{\it eg},
one would like to have an analytic picture of the way the
curvature diverges in the neighborhood of the singularity. By
acquiring such an analytic picture, even if one does hope to
ultimately remedy singularities, one at least wishes to understand
better what is going on near a singularity ({\it eg}, classify
singularities according to their strength \cite{clarke4}) and
perhaps achieve a better control of those unphysical divergences.
To that end, so-called asymptotic growth boundary conditions at
the vicinity of the singularity are usually prescribed
\cite{clarke4}.} exemplifies in our opinion the general {\em
CDG-conservative attitude regarding the appearance and treatment
of singularities in GR}, which we briefly explain now.

Judging by the way we try to define genuine singularities by
elimination and technically (:mathematically) deal with them, on
the one hand, physical spacetime events are identified with the
regular points of $M$---those at which the differential equations
of Einstein hold and they do not suffer from any `differential
geometric disease' ({\it eg}, the differentiability of the
solution metric does not break down in any sense at them as in the
case of the DGSs above). On the other hand, genuine, physically
interesting and significant singularities are pushed---as it were,
by mathematical fiat---to the boundary of the spacetime continuum
in order to preserve the CDG-machinery within the otherwise
regular $M$. This is precisely what we refer to as the {\em
manifold and, {\it in extenso}, CDG-conservatism} in our analysis
of spacetime singularities \cite{clarke4}. Indeed, a
self-referential pun is intended here: {\em the analysis of
spacetime singularities is essentially the (manifold based)
Analysis applied to the study of (true) spacetime singularities,
which, `by definition'}---{\it ie}, by the analytic
inextensibility of $M$ past them---{\em ultimately resist Analysis
(analytic extension)}.

Genuine singularities, as opposed to merely coordinate ones, are
{\it loci} where the manifold based Calculus (Analysis) comes to
an end and hence the manifold based GR is out of its depth ({\it
ie}, the differential equations of Einstein appear to break down
and lose their predictive power---{\it eg}, the solution metrics
blow up and `yield' physically meaningless infinities for
`observable' quantities like the curvature tensor). There is a
tension here: {\em physical} spacetime events, including
coordinate singularities, are the regular points in the interior
$M$ of $\overline{M}$ where CDG applies galore, but {\em physical}
singularities are {\it loci} on $\partial M$) where CDG fails to
apply (breaks down), while, in a paradoxical sense, we seem to
persistently employ CDG (Analytical) means to study the latter
\cite{clarke4}. This CDG-conservatism may be simply understood and
justified on the ground that {\em the only way we so far know how
to do differential geometry is via (the `mediation' in our
calculations---in fact, in our Differential Calculus!---of) a
background continuum space(time), a base differential manifold}.

However, in view of the CDG-monopoly above and with ADG in mind,
we would like to draw a fine line here: while we agree that CDG
(as a {\em mathematical} framework for doing differential
geometry) becomes inadequate at true singularities, we cannot
accept that the {\em physical} law of gravity (modelled after a
{\em differential} equation) breaks down at a singularity all
because we traditionally tend to identify {\em physical} spacetime
with {\em our} mathematical model $M$ which in turn vitally
supports CDG. In this line of thought, Einstein's words from
\cite{einst3} immediately spring to mind:

\begin{quotation}
\noindent ``{\small ...A field theory is not yet completely
determined by the system of field equations. Should one admit the
appearance of singularities?... It is my opinion that
singularities must be excluded. {\em It does not seem reasonable
to me to introduce into a continuum theory points (or lines {\it
etc.}) for which the field equations do not hold}\footnote{Our
emphasis.}...}''
\end{quotation}

\noindent Of course, the distinction we drew above would be simply
unfounded had we not have in our hands not only an alternative (to
CDG) theoretical framework for doing differential geometry
independently of a background $M$, but also had we not been able
within this new framework to formulate Einstein's equations as
{\em differential} equations proper which, {\it a fortiori}, could
be explicitly shown not to be impeded at all by (let alone break
down in) the presence of singularities. ADG is that theoretical
framework.

In connection with the above, a key observation in ADG is that,
since as noted before a differential manifold is nothing else but
the algebra $\smooth(M)$ (or the structure sheaf
$\struc\equiv\smooth_{M}$) of smooth coordinate functions on it
(Gel'fand duality), and since all the singularities in the CDG and
manifold based GR are singularities of some smooth function on
$M$, GR ``{\em carries the seeds of its own destruction}''
\cite{berg0} in the form of singularities exactly because the
physical laws that define it\footnote{We tacitly assume that {\em
a physical theory is defined by the physical laws (:dynamical
equations) formulated within the mathematical framework adopted by
(and adapted to!) that theory}. As noted before, in the case of GR
as originally formulated by Einstein, that mathematical framework
was the CDG and manifold based (pseudo-)Riemannian geometry.} are
mathematically represented by differential equations within the
confines of the CDG-framework. What amounts to the same, the
apparent `self-destructive' feature of GR corresponding to smooth
spacetime singularities is exactly due to the fact that {\em we}
have {\it a priori} posited that {\em physical spacetime} is a
differential manifold.

The crux of the argument here is that it is not the gravitational
field and the law that it obeys that halt or even break down at a
singularity as if they carry the seeds of their own
inapplicability and downfall, but that it is precisely {\em our}
mathematical means of effectuating (representing) that
gravitational dynamics differential geometrically via the
$M$-based CDG---{\it ie}, via $\smooth_{M}$ carrying the germs of
all smooth singularities---that mislead us into thinking that the
CDG-based GR predicts its own autocatastrophe.\footnote{And let it
be stressed here that, in a `Popperian falsifiability' sense, this
is more often than not regarded as a virtue of GR.} And all this
again because we persistently identify {\em physical} spacetime
with a background locally Euclidean space. In other words, it is
the {\em mathematical} notion of a `{\em base manifold}'
(CDG)---or equivalently, {\em our} choice of $\smooth_{M}$ for
structure sheaf of coordinates---in the expression `{\em manifold
based GR}' that carries the differential geometric anomalies in
the guise of singularities that assail GR, and {\em not} the {\em
physical} concept of gravitational field or even the differential
equation that it obeys. Alas, the aforesaid CDG-monopoly and
associated conservatism has misled us into branding $M$ as `{\em
physical spacetime}' and concomitantly made us (con)fuse {\em our}
mathematical framework (CDG) with the {\em physical} theory itself
(GR, gravity) to the extent that we coin genuine singularities as
being {\em physical} ones.

Furthermore, here one could go as far as to maintain that `{\em
Nature has singularities}'. Precisely this we find hard to
swallow: genuine singularities simply pronounce that the
differential manifold based CDG has ceased to be a good
mathematical means (:language) for describing the physical law of
gravity, and that therefore, an alternative mathematical framework
(for doing {\em differential} geometry---provided of course that
the physicist still wishes to represent physical laws by {\em
differential} equations proper) must be sought after. What is
implicit here is our general working philosophy that {\em whenever
there appears to be a discord or asymphony between the mathematics
and the physics, one should invariably question and try to modify
the former, not the latter}. One should blame it on {\em our}
maths, not on {\em Physis} \cite{malrap3,malrap4}.

\paragraph{The ADG-theoretic finitary-algebraic `resolution' of the inner Schwarzschild singularity: a `static', spatial, localized
point-resolution.} This is the neuralgic part of the present paper
in which everything that we have been saying earlier
synergistically comes to effect. We hereby present the
finitistic-algebraic evasion of the interior Schwarzschild
singularity---regarded as a `static', spatial, localized
point-singularity---by ADG-theoretic means in the form of an
outline of the steps of a `syllogism' leading directly to that
`resolution', as follows:

\begin{itemize}

\item First we consider an open and bounded region $X$ of a
spacetime manifold $M$, from which initially, {\it \`a la} Sorkin
\cite{sork0}, we retain only its topological ({\it ie},
$\cont$-continuous) structure---that is, without {\it a priori}
alluding to its differential ({\it ie}, $\smooth$-smooth)
structure.

\item We then let a point-particle of mass $m$ be situated at the `center' of
$X$, as in \cite{df}. That is, we assume that $m$ is a point in
$X$'s interior without evoking any boundary $\partial X$
construction.

\item Next, we cover $X$ by a locally finite open covering $\gauge_{i}$. In
the jargon of ADG, the $U$s in $\gauge_{i}$ are called `{\em open
local gauges}' \cite{mall1,mall2,mall3,malrap1,malrap2,malrap3}.

\item Subsequently, we first discretize $X$ relative to $\gauge_{i}$ in the manner of Sorkin
(\ref{eq13}), and then pass to the Gel'fand-dual representation of
the resulting finitary posets $P_{i}$ in terms of discrete
differential incidence algebras $\omg_{i}$ (\ref{eq16}).

\item Then we consider finsheaves \cite{rap2} of incidence algebras $\Omg_{i}$ in the manner first
introduced in \cite{malrap1}. Parenthetically, one may wish to
bring forth from \cite{rap1} the causet and qauset interpretation
that the $P_{i}$s and their associated $\omg_{i}$s may be given,
as well as the finsheaves thereof \cite{malrap1}.

\item We then recall from (\ref{eq18}) the finitary differential
triads $\triad_{i}$ (of qausets) that the said finsheaves define.
Here the reader may like to remind herself from section 3 of the
two different ways in which we obtained $\triad_{i}$ from $X$. The
first is the step-wise, `{\em constructive}' way, starting from
$P_{i}$ and proceeding via the $\omg_{i}$s and the finsheaves
$\Omg_{i}$ thereof. The second is the `{\em immediate}' way, via
Papatriantafillou's categorical results, going directly from $X$
(now regarded not just as a {\em topological}, but as a {\em
differential} manifold) and the CCDT $\triad_{\infty}$ that it
supports, to $\triad_{i}$, again starting from ({\it ie}, with
base topological spaces) Sorkin's $P_{i}$s. Then one recalls from
(\ref{eq19}) that on these triads the vacuum Einstein equations of
an $fcq$-version of Lorentzian vacuum Einstein gravity hold.

\item Next, from section 4 we recall that the said
finitary differential triads comprise an inverse/direct system
$\invtriad$ possessing, following Sorkin via Papatriantafillou's
categorical perspective on ADG, the CCDT
$\triad_{\infty}\equiv\striad$ as a projective/inductive limit
(\ref{eq20}).

\item Moreover, a plethora of finitary ADG-theoretic
constructions, vital for the formulation of a finitary version of
Lorentzian gravity regarded as a gauge theory, are based on those
$\triad_{i}$s. These include for example the aforementioned
$fcq$-vacuum Einstein equations, the {\em $fcq$-Einstein-Hilbert
action functional} $\eh_{i}$ from which these equations derive
from variation with respect to the Lorentzian gravitational
$fcq$-connections $\conn_{i}$, and the {\em $fcq$-moduli spaces}
$\sconn_{i}(\modl_{i})/\aut\modl_{i}$ of those gauge-equivalent
(self-dual) $fcq$-spin-Lorentzian connections---as noted earlier,
the gauge-theoretic configuration spaces of our $fcq$-version of
Lorentzian (vacuum) Einstein gravity. Thus, it is fitting at this
point to recall from \cite{malrap3}\footnote{Expression (150)
there.} the ``{\em 11-storeys' tower of $fcq$-inverse and direct
systems}'' based on the $\triad_{i}$s in $\invtriad$:\footnote{In
the table below, the letter `$v$' adjoined to our acronym `$fcq$'
stands for `($v$)acuum' \cite{malrap3}. Also, the reader can refer
to the latter paper (or of course to the `originals'
\cite{mall1,mall2,mall3}) for the important notion of `{\em
curvature space}', which however we will not be needing here.}

\vskip 0.2in

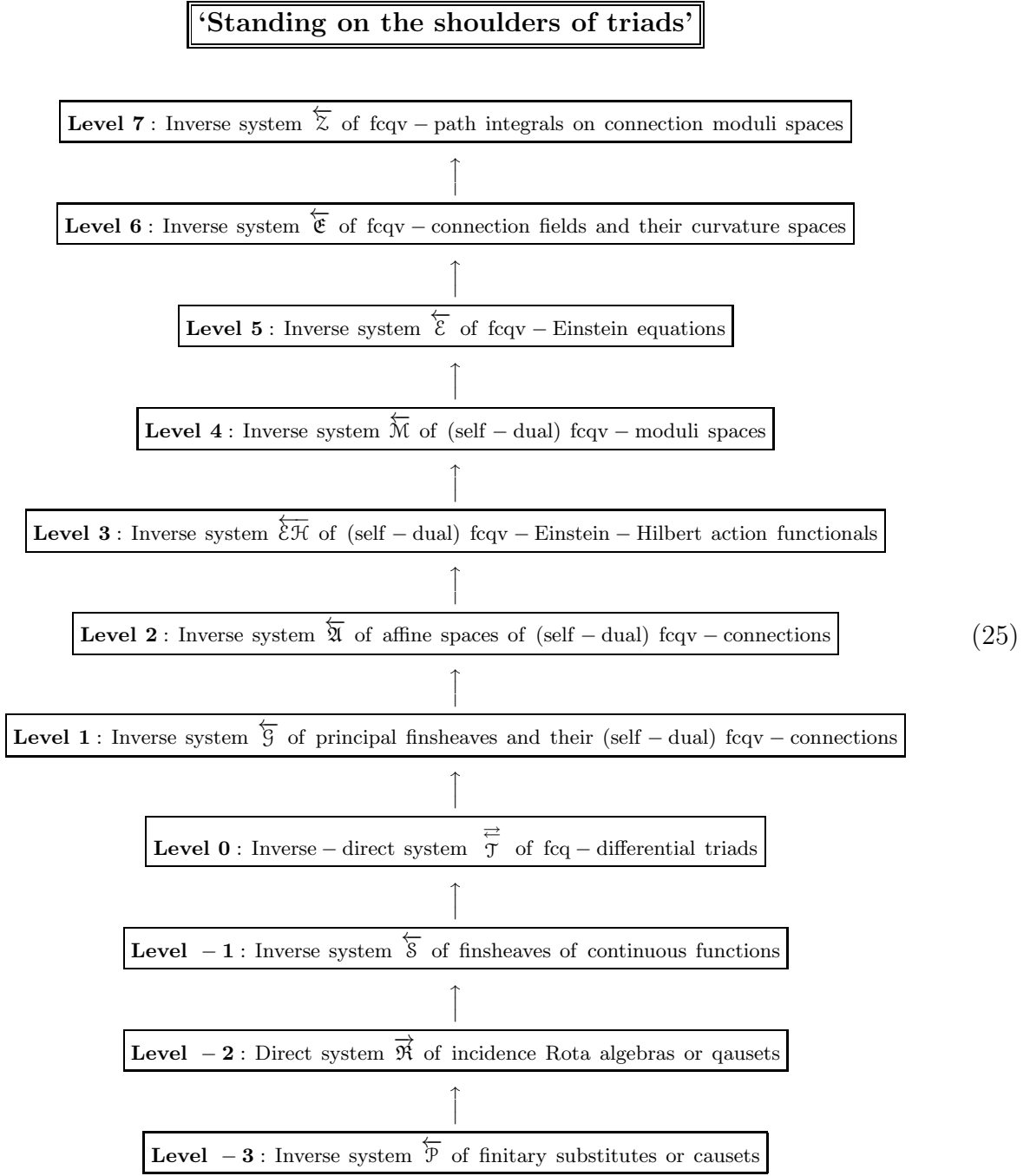
\centerline{\doublebox{\bf `Standing on the shoulders of triads'}}

\medskip

\begin{equation}\label{eq27}
{\fontsize{0.13in}{0.13in}
\begin{CD}
\boxed{\mathbf{Level~7:}~\mathrm{Inverse~system~\invcq~of~fcqv-path~integrals~on~connection~moduli~spaces}}\\
@AAA\\
\boxed{\mathbf{Level~6:}~\mathrm{Inverse~system~\invel~of~fcqv-connection~fields~and~their~curvature~spaces}}\\
@AAA\\
\boxed{\mathbf{Level~5:}~\mathrm{Inverse~system~\inveinst~of~fcqv-Einstein~equations}}\\
@AAA\\
\boxed{\mathbf{Level~4:}~\mathrm{Inverse~system~\invmod~of~(self-dual)~fcqv-moduli~spaces}}\\
@AAA\\
\boxed{\mathbf{Level~3:}~\mathrm{Inverse~system~\inveh~of~(self-dual)~fcqv-Einstein-Hilbert~action~functionals}}\\
@AAA\\
\boxed{\mathbf{Level~2:}~\mathrm{Inverse~system~\invsconn~of~affine~spaces~of~(self-dual)~fcqv-connections}}\\
@AAA\\
\boxed{\mathbf{Level~1:}~\mathrm{Inverse~system~\finv~of~principal~finsheaves~and~their~(self-dual)~fcqv-connections}}\\
@AAA\\
\boxed{\mathbf{Level~0:}~\mathrm{Inverse-direct~system~\invtriad~of~fcq-differential~triads}}\\
@AAA\\
\boxed{\mathbf{Level~-1:}~\mathrm{Inverse~system~\invs~of~finsheaves~of~continuous~functions}}\\
@AAA\\
\boxed{\mathbf{Level~-2:}~\mathrm{Direct~system~\diromg~of~incidence~Rota~algebras~or~qausets}}\\
@AAA\\
\boxed{\mathbf{Level~-3:}~\mathrm{Inverse~system~\inv~of~finitary~substitutes~or~causets}}
\end{CD}}
\end{equation}

\noindent Papatriantafillou's results secure that all these
inverse-direct systems yield, like Sorkin's original projective
system $\inv$, their classical continuum counterparts at the limit
of infinite resolution of the (base) $P_{i}$s. Equivalently, the
continuum structures arise at the limit of infinite (topological)
$\gauge_{i}$-refinement \cite{sork0} (or equivalently, at the
limit of infinite sheaf-theoretic localization of
qausets---inhabiting the stalks of the respective finsheaves at
the finitary level---over $X$'s points).

\item Of special interest to the proposed `resolution' of the
interior Schwarzschild singularity here, is the inverse system
$\inveinst$ at level 5 in (\ref{eq27}) above. {\em The projective
limit of this system recovers the classical continuum vacuum
Einstein equations over the whole ({\it ie}, over all the points
of) $X$} (\ref{eq21}). In particular, we wish to emphasize that

\begin{quotation}
\noindent {\em the (vacuum) Einstein equations hold over the
genuinely singular from the CDG-theoretic vantage point-mass $m$
in the interior of $X$, and in no sense}---at least in the
differential geometric sense of the DGSs in which we are
especially interested in the present paper---{\em do they appear
to break down there}.
\end{quotation}

\noindent In this sense we say that the inner Schwarzschild
singularity has been `resolved' by finitary-algebraic
ADG-theoretic means.

\end{itemize}

Below, we wish to make some further points in order to qualify
more this `resolution':

\begin{itemize}

\item First, as noted in section 2, since in the ADG-theoretic
perspective on GR it is the algebraic $\struc$-connection $\conn$
and {\em not} the smooth metric $g$ (as in the original
formulation of the theory) that is the sole, fundamental
(dynamical) variable, and since moreover ADG is genuinely smooth
background manifold independent, the usual conception of the inner
Schwarzschild singularity as a DGS is {\em not} valid in our
scheme since neither the metric nor its $C^{k}$-extensibility
($k=0\ldots\omega$) are relevant, let alone important, issues in
the theory.

\item Related to the point above is the fact that in ADG we
replace the usual CDG-based GR conception of a genuinely
non-singular spacetime `{\em the solution metric holds ({\it ie},
it is non-singular) in the entire manifold $X$}' by the expression
that `{\em the field law ({\it ie}, the differential equation of
Einstein that $\conn$ defines via its curvature $\ricci$) is valid
throughout all the field's carrier (sheaf) space $\modl$ over the
whole base topological space(time) $X$, which functional sheaf can
in turn host all kinds of singularities}'. {\it Alias}, there is
no breakdown whatsoever of `differentiability', that is, of the
differential equation that $\conn$ defines, in our scheme. The
ADG-gravitational field field $(\conn ,\modl)$, and the dynamical
differential equations that it defines via its curvature,
$\ricci(\conn)(\modl)=0$, is not impeded at all by any
singularities that the background topological space $X$ (or the
functional sheaf $\modl$ localized on it) might possess.

\item One should note that the particular finitary-algebraic inner
Schwarzschild singularity `resolution' presented above is closely
akin to (or one might even say that it follows suit from) the
topological resolution of $X$ {\it \`a la} Sorkin \cite{sork0}, in
the following sense: as the ur-cell $\Lambda(m)|_{\gauge_{i}}$
blowing up and smearing the classically offensive point $m\in X$
becomes `smaller' and `smaller' with topological
$\gauge_{i}$-refinement (resp., the topology $\tau_{i}$ generated
by the open sets in the $\gauge_{i}$s becomes finer and finer),
the law of gravity holds as close to the point-singularity $m$ as
one wishes to get ({\it ie}, at every level `$i$' of resolution or
topological refinement of $X$ by the open coverings $\gauge_{i}$).
Furthermore, at the (projective) limit of infinite topological
resolution (refinement) of $X$ into its points, one gets that
(\ref{eq21}) actually holds on (over) $m$ itself.

\item In connection with the last remarks, it is also worth
pointing out that the law of gravity holds {\em both} at the
`discrete', $fcq$-level of the $P_{i}$s ($\forall i$) {\em and} at
the classical level limit corresponding to $X$, which further
supports our {\it motto} that the ADG-picture of (vacuum)
Lorentzian Einstein gravity (GR), and the $fcq$-version of it, is
genuinely background independent---{\it ie}, whether that
background is a `classical continuum' or a `quantal discretum'.
{\it In toto}, this emphasizes that our ADG-perspective on gravity
is manifestly (base) spacetime free
\cite{malrap1,malrap2,malrap3,malrap4}. With respect to the
CDG-problem of the inner Schwarzschild singularity and the usual
divergence of the gravitational field strength ($\ricci$) in its
vicinity, this freedom may be interpreted as follows: the vacuum
Einstein equations hold {\em both} when a (locally) finite {\em
and} an uncountable continuous infinity of degrees of freedom of
the gravitational field are excited (as it were, when the
gravitational field `occupies' and effectuates a finite and an
infinite number of point-events in the background space(time)
$X$). Moreover, unlike the CDG-based picture of inner
Schwarzschild singularity, no infinity at all (in the analytical
sense of CDG)\footnote{For example, when $m$ is relegated to $X$'s
boundary $\partial X$ and a suitable topology is given to
$\overline{X}=X\cup\partial X$, as $\ricci\mapto m$, $\ricci$
diverges as $1/r^{6}$.} for $\ricci$ is involved as $m$ is
`approached' (in the categorical limit sense of $\infty\leftarrow
i$) by $\ricci_{i}(\conn_{i})$ upon (topological) refinement of
the $\Lambda(m)|_{\gauge_{i}}$s. There is no unphysical infinity
associated with this ADG-picture of the inner Schwarzschild
singularity, and in this sense the latter is `resolved into
locally finite effects'.

\item Of course, all this can be attributed to the fact that the
base topological space(time) $X$ (whether a continuum or a
discontinuum) plays no role whatsoever in the inherently algebraic
differential geometric mechanism of ADG, which, as noted earlier,
derives from the algebra inhabited stalks of the (fin)sheaves
involved and not from the base space which is merely a topological
space. Technically speaking, this is reflected by the fact that
the categorical in nature ADG-formulation of the relevant
differential equations (here, the Einstein equations) involves
(equations between) {\em sheaf morphisms}, and in particular,
$\struc$-morphisms such as $\ricci$. Sheaf morphisms by definition
`see through' the arbitrary base topological space $X$, which in
turn serves only as a surrogate scaffolding, having no physical
significance whatsoever as it plays no role in the gravitational
dynamics---the (vacuum) Einstein differential equations
(\ref{eq9}). $X$ is used only for the mathematical
(:sheaf-theoretic) localization and concomitant gauging of the
algebraic objects in the relevant (fin)sheaves \cite{malrap1}.

\item Even more important than the remarks about the physical
insignificance of the base space $X$, but closely related to them,
is the issue of the $\struc$-functoriality of dynamics already
alluded to in section 2. Namely, the fact that the vacuum Einstein
equations (\ref{eq9}) are (local) expressions of the curvature
$\ricci$ of the gravitational connection $\conn$, which curvature
is an $\struc$-morphism (or $\struc$-tensor)---a `geometrical
object' in ADG jargon \cite{malrap3}, means that our generalized
coordinates (or `measurements') in the structure sheaf $\struc$
(that {\em we} assume to coordinatize the gravitational field
$\conn$ and solder it on $\modl$, which is anyway locally
$\struc^{n}$) respect the gravitational field (strength).
Equivalently, it indicates that the field dynamics `sees through'
our (local) measurements in $\struc(U)$. In turn, since all the
singularities are inherent in $\struc$---the structure sheaf of
generalized algebras of `differentiable' coordinate functions, it
follows that the $\struc$-functorial field dynamics `sees through'
the singularities built into the $\struc$ that we assume.
Equivalently, but in a more philological sense, $\struc$ (and the
singularities that it carries) is `transparent' to the
$\ricci(\conn)$ engaging into the gravitational field
dynamics---the differential equations of Einstein (\ref{eq9}).
{\em In summa}, the field $(\modl ,\conn)$ (and the differential
equation that it defines via its curvature) does not stumble on or
break down at any singularity inherent in $\struc$, since it
passes `through' (or over) them. In this sense, the term
`singularity-resolution' is not a very accurate name to describe
how ADG evades singularities. Perhaps a better term is `{\em
dissolution}' or `{\em absorption}' in $\struc$.

\end{itemize}

\noindent A good example of the aforesaid singularity-dissolution
or absorption in $\struc$ is the ADG-theoretic evasion of the
inner Schwarzschild singularity regarded now as a time-extended
(distributional) {\em spacetime foam dense singularity} in the
sense of Mallios and Rosinger
\cite{malros1,malros2,mall3,mall9,malros3}. We briefly discuss
this `dissolution' next, leaving a thorough treatment to the
forthcoming `paper-book' \cite{malrap4}.

\paragraph{A second `resolution' of the inner Schwarzschild singularity via spacetime foam dense
singularities: a `temporal', distributional time-line resolution.}
There is another possible evasion of the interior Schwarzschild
singularity by ADG-theoretic and finitary means, by regarding it
this time not as a `static' (stationary), `spatial',
point-localized singularity as above, but as an extended,
distributional one (much in the sense of SFSs above) extending
along the `wristwatch' (locally) Euclidean time-axis $L_{t}$ of
the point-particle (\ref{eq22}).

The idea is to regard $L_{t}$ as being inhabited by so-called
`{\em spacetime foam dense singularities}' {\it \`a la}
\cite{malros2,malros3}. On the side, in mathematics these are
singularities of generalized functions (distributions)---situated
on dense subsets of finite-dimensional Euclidean and locally
Euclidean space(time)s (manifolds)---functions which have been
used as coefficients in and have been occurring as solutions of
non-linear (both hyperbolic and elliptic) partial differential
equations, as originally discovered and subsequently developed
entirely algebraically by Rosinger in a series of papers
\cite{ros1,ros2,ros3,ros}. {\it En passant}, these distributions
can be organized into differential algebras generalizing (and
including) both the usual smooth functions $\smooth(M)$ on
manifolds and the well known linear distributions of Schwartz.
They form the basis of Rosinger's non-linear distribution theory.
In physics, interest in such singularities has arisen recently in
the study of `spacetime foam' structures in GR and QG, as studied
primarily by the Polish school of Heller {\it et al.} in the
context of generalized differential spaces
\cite{hell0,hell2,hell3,hell5,hell6,hell1,hell}.

In the context of (applications of) ADG, the said algebras have
been organized into sheaves and used as structure sheaves in the
theory, replacing and generalizing (actually, containing!) the
classical one $\smooth_{X}$. Indeed, classical (CDG) constructions
and results, normally based on $\smooth_{M}$ over a differential
manifold $M$ ({\it eg}, de Rham's theorem, Poincar\'{e}'s lemma,
de Rham cohomology, Weil's integrality theorem, the Chern-Weil
theorem {\it etc}), carry through, virtually unaltered, to the
`ultra-singular' realm of the spacetime foam dense singularities
of the said generalized functions \cite{malros2,malros3,mall9};
moreover, the vacuum Einstein equations are seen again to hold, in
full force, in their very presence \cite{mall3,mall9}.

To comment a bit on the dense singularities, they are arguably the
most robust and numerous singularities that have appeared so far
in the general theory of non-linear partial differential
equations, but three of their most prominent features that we
would like to highlight here, in comparison to the usual
singularities carried by $\smooth_{X}$, are:

\begin{itemize}

\item First, their cardinality. These are singularities on
arbitrary subsets of the underlying topological space(time) $X$.
In particular, they can be concentrated on {\em dense} subsets of
$X$, under the proviso that their complements, consisting of
non-singular (regular) points, are also dense in $X$. In case $X$
is a Euclidean space or a finite-dimensional manifold, {\em the
cardinality of the set of singular points may be larger than that
of the regular ones}. For instance, when one takes $X=\R$ (as we
intuit to do here with $L_{t}$), the dense singular subsets of it
may have the cardinal of the continuum---{\it ie}, the
singularities are situated on the irrational numbers, while the
regular ones are also dense but countable in $\R$ and situated,
say, on the rationals.

\item Second, their situation in the manifold's {\em bulk}. As it
is evident from the above, the dense singularities, apart from
their uncountable multiplicity, are not situated merely at the
boundary of the underlying (topological) space(time) (manifold),
but occupy `central' points in its `bulk-interior'. This is in
striking contrast to the usual theory of $\smooth$-smooth
spacetime singularities that we briefly revisited in section 2
\cite{hawk0,clarke3,clarke4,rendall}, which we may thus coin
`separated and isolated', or `solitary', or even `{\em spacetime
marginal}' for effect. This situation is also in contrast to the
`algebraically generalized differential spaces' (:spacetime foam)
approach to GR and QG of Heller {\it et al.}, as they too assume
(even though they too tend to employ sheaf-theoretic methods) that
singularities---in fact, merely {\em nowhere dense singularities}
(not dense ones!) in the sense of \cite{malros1}---sit right at
the edge of the spacetime manifold (see
\cite{hell0,hell1,hell2,hell4,hell,hell5,hell3}, and especially
\cite{hell6}).

\item And third, as briefly alluded to above, the differential
algebras of generalized functions in Rosinger's non-linear
distribution theory contain both the usual algebra $\smooth(X)$ of
smooth functions and Schwartz's linear distributions
\cite{malros2,malros3}. Furthermore, these non-linear
distributions, either with nowhere dense, or even more
prominently, with dense singularities, have proven to be more
versatile (and potentially more useful in differential geometric
applications) than the, quite popular lately in the theory of
non-linear PDEs, non-linear Colombeau distributions
\cite{colombeau}.\footnote{See \cite{malros2} for a discussion of
the (differential geometric) virtues of Rosinger's distributions
compared to Colombeau's.}

\end{itemize}

\paragraph{Two alternative distributional ADG-resolutions of the inner Schwarzschild singularity.}
Like in the point-resolution presented above, here too we can
evade by ADG-theoretic means the interior Schwarzschild
singularity, regarded as an extended distributional (:SFS-like)
spacetime foam dense singularity along $L_{t}\simeq\R$, in two
different ways---one `direct', the other `indirect' and along the
`finitary' lines of Sorkin. Let us briefly mention the two
strategies, leaving the rather lengthy technical details for
\cite{malrap4}.

\begin{itemize}

\item {\bf `Direct' distributional evasion:} Here, following
\cite{mall3,mall9}, we can directly employ sheaves of Rosinger's
generalized functions hosting dense singularities on $L_{t}$ as
coordinate structure sheaves in the theory. Then we
straightforwardly borrow the main result from \cite{mall3,mall9},
namely, that Einstein's equations hold over all $L_{t}$ when
Rosinger's spacetime foam sheaves are used as $\struc$. We call
this strategy `direct', because, like in the first
`non-constructive' point-resolution before which evoked
Papatriantafillou's results and straightforwardly defined finitary
differential triads on the $\stackrel{\gauge_{i}}{\sim}$-moduli
spaces $P_{i}$, one can directly define spacetime foam
differential triads without having to go `constructively', in a
roundabout way, via finitary coverings, finsheaves (of incidence
algebras) {\it etc.} The latter we can accomplish in the second
possible strategy briefly described next.

\item {\bf `Indirect' distributional evasion:} Here, we combine
the approach of Mallios-Rosinger in \cite{malros2,malros3} with
Sorkin's in \cite{sork0} and let $X\equiv I\subset L_{t}\simeq\R$
($I$ a bounded interval of the real line as befits a physically
realistic point-particle of finite lifetime) be covered by locally
finite `{\em singularity-open coverings}'. These are covering
families of open subsets of $X$ containing singularities (of
Rosinger's generalized functions) densely at their points. Then we
go to finsheaves (of incidence algebras) and the finitary
differential triads picture thereof so as to show that for each
such covering the vacuum Einstein equations hold {\it \`a la}
\cite{malrap3} and \cite{malrap3}, and finally we pass to the
classical `continuum' projective limit of maximum topological-{\it
cum}-singularity refinement to show that the vacuum Einstein
equations hold over the whole (space)time---in particular, over
all $X$. To be precise, and in keeping with Sorkin's inverse limit
result mentioned earlier, the vacuum Einstein equations can be
seen to hold over all the densely singular points of $X$---itself
assumed to be populated with spacetime foam dense
singularities---when recovered as a dense subset of (closed points
of) the non-Hausdorff inverse limit space of Sorkin's finitary
substitutes and the differential triads they support relative to
the said locally finite open singularity-covers.

\end{itemize}

\section{Epilegomena: Implications for Quantum Gravity}

In this concluding section we remark briefly on two issues. First,
how ADG may prove to be a suitable theoretical framework in which
to formulate a genuinely background independent QG. Also, since
ADG appears to evade completely (gravitational) singularities
\cite{mall3,mall7,mall9,mall11,mall4,malrap4}, we touch in its
light on the nowadays general consensus (or at least, the wishful
expectation) that QG should resolve, or ultimately remove,
spacetime singularities \cite{hawk2,pen5,husain}. In this context,
we loosely compare the evasion of the interior Schwarzschild
singularity presented above with a similar resolution of it
achieved very recently by the methods and results of Loop QG (LQG)
in \cite{modesto}, making some relevant comments in the process.
However, we leave a thorough discussion of what follows to
\cite{malrap4}.

\begin{itemize}

\item {\bf Genuinely background independent QG.} A major issue in
QG, especially in non-perturbative canonical QGR \cite{thiem2} in
its connection based LQG version \cite{rovsm1,thiem3,smolin}, is
to formulate the theory in a genuinely background independent
fashion \cite{alvarez}. In a nutshell, by `{\em background
independence}' it is meant `{\em background metric independence}'.
That is, unlike in the usual (mainly perturbative) approaches to
QG where one fixes a (usually flat, Minkowski) background metric
in order to formulate the quantum dynamics (and expand the
relevant quantities about it, as well as to impose physically
meaningful commutation relations among them), here there is no
such desire since, anyway, it appears to be begging the question
to fix {\it a priori} (and by hand!), and moreover to duplicate,
the supposedly sole dynamical variable of GR---the spacetime
geometry (metric). Ashtekar and coworkers have succeeded over the
years in formulating LQG in a manifestly fixed background metric
independent way \cite{ash5}. Alas, {\em a smooth spacetime
manifold is still retained in the background} \cite{ash4}---or
else, how could one still use differential geometric ideas and
constructions \cite{ashlew2} in QG research? For example, as noted
earlier, the new connection variables \cite{ash} employed in LQG
are {\em smooth} (spin-Lorentzian) connections based on a
differential spacetime manifold, let alone that the smooth metric
is still implicitly present in the guise of the smooth comoving
tetrad (:{\it vierbein}) field variables (1st-order formalism).
All this is another instance of the aforementioned base manifold
and CDG-conservatism and monopoly.

By contrast, in ADG GR is not only formulated, as befits a purely
gauge theory, solely in terms of the gravitational
$\struc$-connection variable without at all the presence of a
metric (`half-order formalism'), but also, {\it a fortiori}, {\em
no base differential spacetime manifold appears at all in the
theory}. In this sense, the ADG-approach to gravity---classical or
quantum---is genuinely background independent
\cite{malrap3,malrap4}. On the other hand, it is plain that since
singularities are inherent in $\smooth_{M}$ ({\it ie}, in the
background differential spacetime manifold $M$), loop QG still has
to reckon with them---that is, they are still problems for the
theory and thus the theory still aims at resolving them somehow.
We thus comment on a recent resolution of the inner Schwarzschild
singularity by LQG means \cite{modesto} in the next paragraph,
comparing it at the same time with ours above.

\item {\bf Comparison with a recent resolution of the inner
Schwarzschild singularity by LQG methods.} As noted before, there
is currently optimism among theoretical physicists that QG will
shed more light and ultimately (re)solve the problem of smooth
spacetime singularities in GR. Notably, within the past three
years, in the context of Loop Quantum Cosmology it has been shown
that the initial (`Big Bang') singularity predicted by GR can be
indeed resolved \cite{boj0,boj2,ash3,husain}. However, even more
remarkable for the present paper is the following very recent
result of Modesto \cite{modesto}, which was also arrived at by LQG
means: in one sentence, {\em the Schwarzschild black hole
singularity of the classical theory (GR) `disappears' in QG}. In
this penultimate paragraph we would like to describe briefly this
`disappearance', comment on it and juxtapose it with the
`resolution' of the same singularity that we achieved herein by
ADG-theoretic means.

Let us first note that since, as mentioned before, LQG, although
background metric independent, still employs a base differential
(spacetime) manifold for its constructions, the problem of
singularities in the classical theory persists and has to be
reckoned with in the quantum theory. In this regard, it is fair to
say that loop QG `expects' that the `true' quantum theory of
gravity it aspires to be should ultimately resolve or remove the
singularities and the associated pathological infinities of the
classical theory \cite{husain}. Briefly, in \cite{modesto} the
interior Schwarzschild singularity appears to be resolved as
follows:\footnote{The reader is referred to \cite{modesto} for
detailed arguments, calculations and pertinent citations.}

\vskip 0.1in

1. To begin with, one expresses the Ricci scalar curvature, which
as noted earlier blows up as $1/r^{6}$ near the interior ($r=0$)
Schwarzschild singularity, in terms of the spacetime volume.

\vskip 0.1in

2. Then, one evokes {\em the} major result in LQG, namely, that
the said volume is quantized---{\it ie}, it is promoted to a
volume operator having a discrete eigen-spectrum. Parenthetically
we mention that this volume-quantization \cite{ashlew4} is just
one of a series of significant results in Ashtekar's quantum
(Riemannian) geometry programme accompanying LQG
\cite{ash4,thiem2,thiem3,smolin}, along with the quantization of
length \cite{thiem1} and area \cite{ashlew3} (see also
\cite{rovsm,rovelli}). Thus, near the Schwarzschild black hole,
$\ricci$ is rendered finite and the classical infinities are
controlled (`regularized') by quantum theory.

\vskip 0.1in

3. Moreover, one can show that the said `regularization' is not
`kinematical' and without physical significance---one that is {\it
a priori} fixed by hand like for example the space(time)
discretizations in lattice QCD---but it is a dynamical one. This
is so because the Hamiltonian (constraint), which regulates the
dynamical time-evolution in the canonical approach to GR
classically underlying LQG \cite{thiem2,thiem3}, can also be
expressed in terms of the volume operator. Thus, as Modesto shows,
{\em the spacetime can be dynamically extended past the interior
Schwarzschild singularity, with no infinity involved at all}.

\vskip 0.1in

4. On the other hand, from a {\em differential} geometric
viewpoint, the upshot of all this is that the said dynamical
evolution, which is classically represented by a {\em
differential} equation on the spacetime continuum,\footnote{After
all, the Hamiltonian (constraint) in the classical canonical
theory (GR) is the generator of time-diffeomorphisms.} is now
substituted, in view of the said quantization of spacetime
geometry in LQG, by a `discrete', {\em difference equation}
(discretely parameterized by the coefficients of the physical
quantum eigenstates of the volume operator---in a quantum
cosmological setting, see also \cite{boj2}). {\it In summa}, one
can say that {\em the inner Schwarzschild singularity is resolved
due to the quantization of spacetime itself}.

\end{itemize}

\noindent Based on the brief description above, our comparison of
the two `resolutions' of the interior Schwarzschild singularity
focuses on two fundamental in our opinion differences:

\vskip 0.1in

{\bf I.} Unlike in the LQG `resolution' where a quantization of
spacetime appears to be necessary, in the ADG `resolution' this is
not so, for the theory is `intrinsically background
spacetimeless'. That is, the theory is indifferent as to whether
that background is a `classical continuum' or a `quantal
discretum', since the dynamical Einstein equations hold both at
the `discrete-quantal' (finitary) level {\em and} at the
`continuous-classical' (infinitary) one \cite{malrap3,malrap4}. In
the ADG perspective on gravity, where the sole dynamical variable
is an algebraic connection field $\conn$ on a vector/agebra sheaf
$\modl$ (on an in principle arbitrary topological space $X$), the
quest for a quantization of spacetime is virtually begging the
question: {\em in the first place, in ADG, what `spacetime' is one
talking about?}. Another way to say this is that, from the
ADG-viewpoint, gravity ({\it ie}, the dynamically autonomous
gravitational field $\conn$ defining the Einstein equations via
its curvature) has nothing to do with a background `space(time)'
(in our case, the background $X$ which serves only as a surrogate
topological space for the sheaf-theoretic localization and
representation of the relevant sheaves; {\it eg}, $\struc$,
$\modl$ and $\conn$ acting on it), so that a possible quantum
theory of the former is in no need of a quantum description of the
latter \cite{malrap2,malrap3,malrap4}. As a consequence of this
difference,

\vskip 0.1in

{\bf II.} Unlike in the loop QG `resolution' where the said
spacetime quantization and concomitant discretization appears to
necessitate the abandoning of the picture of `gravitational
dynamical evolution' as a differential equation proper (and, as a
result, the abandonment of differential geometric ideas in the
quantum regime---{\it eg}, see Isham quotation from \cite{ish}
before), in the ADG `resolution' all the differential geometric
machinery (of the background spacetime continuum) is retained in
full effect \cite{malrap2,malrap3,malrap4}. Moreover, this is so
manifestly independently of that background, and {\it a fortiori},
even if that background is taken to be a `discretum' where
differential geometric ideas would traditionally---{\it ie}, from
the CDG-viewpoint of the continuous manifold---be expected to fail
to apply.

\section*{Acknowledgments}

The author is indebted to Chris Isham for numerous conversations
about the potential import of ADG to classical and quantum
gravity, and of course to Tasos Mallios for orienting, guiding and
advising him about selecting and working out what may prove to be
of importance to classical and quantum gravity research from the
wealth of mathematical physics ideas that ADG is pregnant to. The
present paper is just the tip of an `iceberg' of a recent
`paper-book' \cite{malrap4}, written in collaboration with
Mallios, on a detailed treatment of $\smooth$-smooth gravitational
singularities and their possible evasion by ADG-theoretic means.
This author also wishes to acknowledge financial support from the
European Commission in the form of a European Reintegration Grant
(ERG-CT-505432) held at the University of Athens, Greece.

\end{document}

%% file: diagram.tex
\makeatletter

\def\diagram{\m@th\leftwidth=\z@ \rightwidth=\z@ \topheight=\z@
\botheight=\z@ \setbox\@picbox\hbox\bgroup}

\def\enddiagram{\egroup\wd\@picbox\rightwidth\unitlength
\ht\@picbox\topheight\unitlength \dp\@picbox\botheight\unitlength
\hskip\leftwidth\unitlength\box\@picbox}

\def\bfig{\begin{diagram}}
\def\efig{\end{diagram}}
\newcount\wideness \newcount\leftwidth \newcount\rightwidth
\newcount\highness \newcount\topheight \newcount\botheight

\def\ratchet#1#2{\ifnum#1<#2 \global #1=#2 \fi}

\def\putbox(#1,#2)#3{%
\horsize{\wideness}{#3} \divide\wideness by 2 {\advance\wideness
by #1 \ratchet{\rightwidth}{\wideness}} {\advance\wideness by -#1
\ratchet{\leftwidth}{\wideness}} \vertsize{\highness}{#3}
\divide\highness by 2 {\advance\highness by #2
\ratchet{\topheight}{\highness}} {\advance\highness by -#2
\ratchet{\botheight}{\highness}} \put(#1,#2){\makebox(0,0){$#3$}}}

\def\putlbox(#1,#2)#3{%
\horsize{\wideness}{#3} {\advance\wideness by #1
\ratchet{\rightwidth}{\wideness}} {\ratchet{\leftwidth}{-#1}}
\vertsize{\highness}{#3} \divide\highness by 2 {\advance\highness
by #2 \ratchet{\topheight}{\highness}} {\advance\highness by -#2
\ratchet{\botheight}{\highness}}
\put(#1,#2){\makebox(0,0)[l]{$#3$}}}

\def\putrbox(#1,#2)#3{%
\horsize{\wideness}{#3} {\ratchet{\rightwidth}{#1}}
{\advance\wideness by -#1 \ratchet{\leftwidth}{\wideness}}
\vertsize{\highness}{#3} \divide\highness by 2 {\advance\highness
by #2 \ratchet{\topheight}{\highness}} {\advance\highness by -#2
\ratchet{\botheight}{\highness}}
\put(#1,#2){\makebox(0,0)[r]{$#3$}}}

\def\adjust[#1]{} % For compatibility

\newcount \coefa
\newcount \coefb
\newcount \coefc
\newcount\tempcounta
\newcount\tempcountb
\newcount\tempcountc
\newcount\tempcountd
\newcount\xext
\newcount\yext
\newcount\xoff
\newcount\yoff
\newcount\gap%
\newcount\arrowtypea
\newcount\arrowtypeb
\newcount\arrowtypec
\newcount\arrowtyped
\newcount\arrowtypee
\newcount\height
\newcount\width
\newcount\xpos
\newcount\ypos
\newcount\run
\newcount\rise
\newcount\arrowlength
\newcount\halflength
\newcount\arrowtype
\newdimen\tempdimen
\newdimen\xlen
\newdimen\ylen
\newsavebox{\tempboxa}%
\newsavebox{\tempboxb}%
\newsavebox{\tempboxc}%

\newdimen\w@dth

\def\setw@dth#1#2{\setbox\z@\hbox{\m@th$#1$}\w@dth=\wd\z@
\setbox\@ne\hbox{\m@th$#2$}\ifnum\w@dth<\wd\@ne \w@dth=\wd\@ne \fi
\advance\w@dth by 1.2em}

\def\t@^#1_#2{\allowbreak\def\n@one{#1}\def\n@two{#2}\mathrel
{\setw@dth{#1}{#2} \mathop{\hbox to
\w@dth{\rightarrowfill}}\limits \ifx\n@one\empty\else
^{\box\z@}\fi \ifx\n@two\empty\else _{\box\@ne}\fi}}
\def\t@@^#1{\@ifnextchar_{\t@^{#1}}{\t@^{#1}_{}}}
\def\to{\@ifnextchar^{\t@@}{\t@@^{}}}

\def\t@left^#1_#2{\def\n@one{#1}\def\n@two{#2}\mathrel{\setw@dth{#1}{#2}
\mathop{\hbox to \w@dth{\leftarrowfill}}\limits
\ifx\n@one\empty\else ^{\box\z@}\fi \ifx\n@two\empty\else
_{\box\@ne}\fi}}
\def\t@@left^#1{\@ifnextchar_{\t@left^{#1}}{\t@left^{#1}_{}}}
\def\toleft{\@ifnextchar^{\t@@left}{\t@@left^{}}}

\def\two@^#1_#2{\allowbreak
\def\n@one{#1}\def\n@two{#2}\mathrel{\setw@dth{#1}{#2}
\mathop{\vcenter{\lineskip\z@\baselineskip\z@
                 \hbox to \w@dth{\rightarrowfill}%
                 \hbox to \w@dth{\rightarrowfill}}%
       }\limits
\ifx\n@one\empty\else ^{\box\z@}\fi \ifx\n@two\empty\else
_{\box\@ne}\fi}}
\def\tw@@^#1{\@ifnextchar _{\two@^{#1}}{\two@^{#1}_{}}}
\def\two{\@ifnextchar ^{\tw@@}{\tw@@^{}}}

\def\tofr@^#1_#2{\def\n@one{#1}\def\n@two{#2}\mathrel{\setw@dth{#1}{#2}
\mathop{\vcenter{\hbox to \w@dth{\rightarrowfill}\kern-1.7ex
                 \hbox to \w@dth{\leftarrowfill}}%
       }\limits
\ifx\n@one\empty\else ^{\box\z@}\fi \ifx\n@two\empty\else
_{\box\@ne}\fi}}
\def\t@fr@^#1{\@ifnextchar_ {\tofr@^{#1}}{\tofr@^{#1}_{}}}
\def\tofro{\@ifnextchar^ {\t@fr@}{\t@fr@^{}}}

\def\mon{\mathop{\m@th\hbox to
      14.6\P@{\lasyb\char'51\hskip-2.1\P@$\arrext$\hss
$\mathord\rightarrow$}}\limits} % width of \epi
\def\leftmono{\mathrel{\m@th\hbox to
14.6\P@{$\mathord\leftarrow$\hss$\arrext$\hskip-2.1\P@\lasyb\char'50%
}}\limits} % width of \epi
\mathchardef\arrext="0200       % amr minus for arrow extension (see \into)

\def\settypes(#1,#2,#3){\arrowtypea#1 \arrowtypeb#2 \arrowtypec#3}
\def\settoheight#1#2{\setbox\@tempboxa\hbox{#2}#1\ht\@tempboxa\relax}%
\def\settodepth#1#2{\setbox\@tempboxa\hbox{#2}#1\dp\@tempboxa\relax}%
\def\settokens`#1`#2`#3`#4`{%
     \def\tokena{#1}\def\tokenb{#2}\def\tokenc{#3}\def\tokend{#4}}
\def\setsqparms[#1`#2`#3`#4;#5`#6]{%
\arrowtypea #1 \arrowtypeb #2 \arrowtypec #3 \arrowtyped #4 \width
#5 \height #6 }
\def\setpos(#1,#2){\xpos=#1 \ypos#2}

\def\settriparms[#1`#2`#3;#4]{\settripairparms[#1`#2`#3`1`1;#4]}%

\def\settripairparms[#1`#2`#3`#4`#5;#6]{%
\arrowtypea #1 \arrowtypeb #2 \arrowtypec #3 \arrowtyped #4
\arrowtypee #5 \width #6 \height #6 }

\def\resetparms{\settripairparms[1`1`1`1`1;500]\width 500}%default values%

\resetparms

\def\mvector(#1,#2)#3{%%
\put(0,0){\vector(#1,#2){#3}}%
\put(0,0){\vector(#1,#2){26}}%
}
\def\evector(#1,#2)#3{{%%
\arrowlength #3
\put(0,0){\vector(#1,#2){\arrowlength}}%
\advance \arrowlength by-30
\put(0,0){\vector(#1,#2){\arrowlength}}%
}}

\def\horsize#1#2{%
\settowidth{\tempdimen}{$#2$}%
#1=\tempdimen \divide #1 by\unitlength }

\def\vertsize#1#2{%
\settoheight{\tempdimen}{$#2$}%
#1=\tempdimen
\settodepth{\tempdimen}{$#2$}%
\advance #1 by\tempdimen \divide #1 by\unitlength }

\def\putvector(#1,#2)(#3,#4)#5#6{{%
\ifnum3<\arrowtype \putdashvector(#1,#2)(#3,#4)#5\arrowtype \else
\ifnum\arrowtype<-3 \putdashvector(#1,#2)(#3,#4)#5\arrowtype \else
\xpos=#1 \ypos=#2 \run=#3 \rise=#4 \arrowlength=#5 \ifnum
\arrowtype<0
    \ifnum \run=0
        \advance \ypos by-\arrowlength
    \else
        \tempcounta \arrowlength
        \multiply \tempcounta by\rise
        \divide \tempcounta by\run
        \ifnum\run>0
            \advance \xpos by\arrowlength
            \advance \ypos by\tempcounta
        \else
            \advance \xpos by-\arrowlength
            \advance \ypos by-\tempcounta
        \fi
    \fi
    \multiply \arrowtype by-1
    \multiply \rise by-1
    \multiply \run by-1
\fi \ifcase \arrowtype
\or \put(\xpos,\ypos){\vector(\run,\rise){\arrowlength}}%
\or \put(\xpos,\ypos){\mvector(\run,\rise)\arrowlength}%
\or \put(\xpos,\ypos){\evector(\run,\rise){\arrowlength}}%
\fi\fi\fi }}

\def\putsplitvector(#1,#2)#3#4{%%
\xpos #1 \ypos #2 \arrowtype #4 \halflength #3 \arrowlength #3
\gap 140 \advance \halflength by-\gap \divide \halflength by2
\ifnum\arrowtype>0
   \ifcase \arrowtype
   \or \put(\xpos,\ypos){\line(0,-1){\halflength}}%
       \advance\ypos by-\halflength
       \advance\ypos by-\gap
       \put(\xpos,\ypos){\vector(0,-1){\halflength}}%
   \or \put(\xpos,\ypos){\line(0,-1)\halflength}%
       \put(\xpos,\ypos){\vector(0,-1)3}%
       \advance\ypos by-\halflength
       \advance\ypos by-\gap
       \put(\xpos,\ypos){\vector(0,-1){\halflength}}%
   \or \put(\xpos,\ypos){\line(0,-1)\halflength}%
       \advance\ypos by-\halflength
       \advance\ypos by-\gap
       \put(\xpos,\ypos){\evector(0,-1){\halflength}}%
   \fi
\else \arrowtype=-\arrowtype
   \ifcase\arrowtype
   \or \advance \ypos by-\arrowlength
       \put(\xpos,\ypos){\line(0,1){\halflength}}%
       \advance\ypos by\halflength
       \advance\ypos by\gap
       \put(\xpos,\ypos){\vector(0,1){\halflength}}%
   \or \advance \ypos by-\arrowlength
       \put(\xpos,\ypos){\line(0,1)\halflength}%
       \put(\xpos,\ypos){\vector(0,1)3}%
       \advance\ypos by\halflength
       \advance\ypos by\gap
       \put(\xpos,\ypos){\vector(0,1){\halflength}}%
   \or \advance \ypos by-\arrowlength
       \put(\xpos,\ypos){\line(0,1)\halflength}%
       \advance\ypos by\halflength
       \advance\ypos by\gap
       \put(\xpos,\ypos){\evector(0,1){\halflength}}%
   \fi
\fi }

\def\putmorphism(#1)(#2,#3)[#4`#5`#6]#7#8#9{{%
\run #2 \rise #3 \ifnum\rise=0
  \puthmorphism(#1)[#4`#5`#6]{#7}{#8}#9%
\else\ifnum\run=0
  \putvmorphism(#1)[#4`#5`#6]{#7}{#8}#9%
\else
\setpos(#1)%
\arrowlength #7 \arrowtype #8 \ifnum\run=0 \else\ifnum\rise=0
\else \ifnum\run>0
    \coefa=1
\else
   \coefa=-1
\fi \ifnum\arrowtype>0
   \coefb=0
   \coefc=-1
\else
   \coefb=\coefa
   \coefc=1
   \arrowtype=-\arrowtype
\fi \width=2 \multiply \width by\run \divide \width by\rise \ifnum
\width<0  \width=-\width\fi \advance\width by60 \if l#9
\width=-\width\fi
\putbox(\xpos,\ypos){#4}%            %node 1
{\multiply \coefa by\arrowlength%      %node 2
\advance\xpos by\coefa \multiply \coefa by\rise \divide \coefa
by\run \advance \ypos by\coefa
\putbox(\xpos,\ypos){#5} }%
{\multiply \coefa by\arrowlength%      %label
\divide \coefa by2 \advance \xpos by\coefa \advance \xpos by\width
\multiply \coefa by\rise \divide \coefa by\run \advance \ypos
by\coefa
\if l#9%
   \putrbox(\xpos,\ypos){#6}%
\else\if r#9%
   \putlbox(\xpos,\ypos){#6}%
\fi\fi }%
{\multiply \rise by-\coefc%             %arrow
\multiply \run by-\coefc \multiply \coefb by\arrowlength \advance
\xpos by\coefb \multiply \coefb by\rise \divide \coefb by\run
\advance \ypos by\coefb \multiply \coefc by70 \advance \ypos
by\coefc \multiply \coefc by\run \divide \coefc by\rise \advance
\xpos by\coefc \multiply \coefa by140 \multiply \coefa by\run
\divide \coefa by\rise \advance \arrowlength by\coefa
\ifcase\arrowtype
\or \put(\xpos,\ypos){\vector(\run,\rise){\arrowlength}}%
\or \put(\xpos,\ypos){\mvector(\run,\rise){\arrowlength}}%
\or \put(\xpos,\ypos){\evector(\run,\rise){\arrowlength}}%
\fi}\fi\fi\fi\fi}}

\newcount\numbdashes \newcount\lengthdash \newcount\increment

\def\howmanydashes{% Actually returns both number and length
\numbdashes=\arrowlength \lengthdash=40 \divide\numbdashes by
\lengthdash \lengthdash=\arrowlength \divide\lengthdash by
\numbdashes
\increment=\lengthdash \multiply\lengthdash by 3
\divide\lengthdash by 5 }

\def\putdashvector(#1)(#2,#3)#4#5{%
\ifnum#3=0 \putdashhvector(#1){#4}#5 \else \ifnum#2=0
\putdashvvector(#1){#4}#5\fi\fi}

\def\putdashhvector(#1,#2)#3#4{{%
\arrowlength=#3 \howmanydashes
\multiput(#1,#2)(\increment,0){\numbdashes}%
{\vrule height .4pt width \lengthdash\unitlength} \arrowtype=#4
\xpos=#1 \ifnum\arrowtype<0 \advance\arrowtype by 7 \fi
\ifcase\arrowtype \or \advance\xpos by 10
    \put(\xpos,#2){\vector(-1,0){\lengthdash}}
    \advance\xpos by 40
    \put(\xpos,#2){\vector(-1,0){\lengthdash}}
\or \advance \xpos by 10
    \put(\xpos,#2){\vector(-1,0){\lengthdash}}
    \advance\xpos by  \arrowlength
    \advance\xpos by  -50
    \put(\xpos,#2){\vector(-1,0){\lengthdash}}
\or \advance\xpos by 10
    \put(\xpos,#2){\vector(-1,0){\lengthdash}}
\or \advance\xpos by \arrowlength
    \advance\xpos by -\lengthdash
    \put(\xpos,#2){\vector(1,0){\lengthdash}}
\or {\advance\xpos by 10
    \put(\xpos,#2){\vector(1,0){\lengthdash}}}
    \advance\xpos by \arrowlength
    \advance\xpos by -\lengthdash
    \put(\xpos,#2){\vector(1,0){\lengthdash}}
\or \advance\xpos by \arrowlength
    \advance\xpos by -\lengthdash
    \put(\xpos,#2){\vector(1,0){\lengthdash}}
    \advance\xpos by -40
    \put(\xpos,#2){\vector(1,0){\lengthdash}}
   \fi
}}

\def\putdashvvector(#1,#2)#3#4{{%
\arrowlength=#3 \howmanydashes \ypos=#2 \advance\ypos by
-\arrowlength
\multiput(#1,#2)(0,\increment){\numbdashes}%
    {\vrule width .4pt height \lengthdash\unitlength}
\arrowtype=#4 \ypos=#2 \ifnum\arrowtype<0 \advance\arrowtype by 7
\fi \ifcase\arrowtype \or \advance\ypos by \arrowlength
\advance\ypos by -40
    \put(#1,\ypos){\vector(0,1){\lengthdash}}
    \advance\ypos by -40
    \put(#1,\ypos){\vector(0,1){\lengthdash}}
\or \advance\ypos by 10
    \put(#1,\ypos){\vector(0,1){\lengthdash}}
    \advance\ypos by \arrowlength \advance\ypos by -40
    \put(#1,\ypos){\vector(0,1){\lengthdash}}
\or \advance\ypos by \arrowlength \advance\ypos by -40
    \put(#1,\ypos){\vector(0,1){\lengthdash}}
\or \advance\ypos by 10
    \put(#1,\ypos){\vector(0,-1){\lengthdash}}
\or \advance\ypos by 10
    \put(#1,\ypos){\vector(0,-1){\lengthdash}}
    \advance\ypos by \arrowlength \advance\ypos by -40
    \put(#1,\ypos){\vector(0,-1){\lengthdash}}
\or \advance\ypos by 10
    \put(#1,\ypos){\vector(0,-1){\lengthdash}}
    \advance\ypos by 40
    \put(#1,\ypos){\vector(0,-1){\lengthdash}}
\fi }}

\def\puthmorphism(#1,#2)[#3`#4`#5]#6#7#8{{%
\xpos #1 \ypos #2 \width #6 \arrowlength #6 \arrowtype=#7
\putbox(\xpos,\ypos){#3\vphantom{#4}}%
{\advance \xpos by\arrowlength
\putbox(\xpos,\ypos){\vphantom{#3}#4}}%
\horsize{\tempcounta}{#3}%
\horsize{\tempcountb}{#4}%
\divide \tempcounta by2 \divide \tempcountb by2 \advance
\tempcounta by30 \advance \tempcountb by30 \advance \xpos
by\tempcounta \advance \arrowlength by-\tempcounta \advance
\arrowlength by-\tempcountb
\putvector(\xpos,\ypos)(1,0)\arrowlength\arrowtype \divide
\arrowlength by2 \advance \xpos by\arrowlength
\vertsize{\tempcounta}{#5}%
\divide\tempcounta by2 \advance \tempcounta by20
\if a#8 %
   \advance \ypos by\tempcounta
   \putbox(\xpos,\ypos){#5}%
\else
   \advance \ypos by-\tempcounta
   \putbox(\xpos,\ypos){#5}%
\fi}}

\def\putvmorphism(#1,#2)[#3`#4`#5]#6#7#8{{%
\xpos #1 \ypos #2 \arrowlength #6 \arrowtype #7
\settowidth{\xlen}{$#5$}%
\putbox(\xpos,\ypos){#3}%
{\advance \ypos by-\arrowlength
\putbox(\xpos,\ypos){#4}}%
{\advance\arrowlength by-140 \advance \ypos by-70 \ifdim\xlen>0pt
   \if m#8%
      \putsplitvector(\xpos,\ypos)\arrowlength\arrowtype
   \else
   \putvector(\xpos,\ypos)(0,-1)\arrowlength\arrowtype
   \fi
\else
   \putvector(\xpos,\ypos)(0,-1)\arrowlength\arrowtype
\fi}%
\ifdim\xlen>0pt
   \divide \arrowlength by2
   \advance\ypos by-\arrowlength
   \if l#8%
      \advance \xpos by-40
      \putrbox(\xpos,\ypos){#5}%
   \else\if r#8%
      \advance \xpos by40
      \putlbox(\xpos,\ypos){#5}%
   \else
      \putbox(\xpos,\ypos){#5}%
   \fi\fi
\fi }}

\def\putsquarep<#1>(#2)[#3;#4`#5`#6`#7]{{%
\setsqparms[#1]%
\setpos(#2)%
\settokens`#3`%
\puthmorphism(\xpos,\ypos)[\tokenc`\tokend`{#7}]{\width}{\arrowtyped}b%
\advance\ypos by \height
\puthmorphism(\xpos,\ypos)[\tokena`\tokenb`{#4}]{\width}{\arrowtypea}a%
\putvmorphism(\xpos,\ypos)[``{#5}]{\height}{\arrowtypeb}l%
\advance\xpos by \width
\putvmorphism(\xpos,\ypos)[``{#6}]{\height}{\arrowtypec}r%
}}

\def\putsquare{\@ifnextchar <{\putsquarep}{\putsquarep%
   <\arrowtypea`\arrowtypeb`\arrowtypec`\arrowtyped;\width`\height>}}
\def\square{\@ifnextchar< {\squarep}{\squarep
   <\arrowtypea`\arrowtypeb`\arrowtypec`\arrowtyped;\width`\height>}}
\def\squarep<#1>[#2`#3`#4`#5;#6`#7`#8`#9]{{%       %     #2------>#3
\setsqparms[#1]%                                   %      |       |
\diagram%                                          %      |       |
\putsquarep<\arrowtypea`\arrowtypeb`\arrowtypec`%  %    #7|       |#8
\arrowtyped;\width`\height>%                       %      |       |
(0,0)[#2`#3`#4`{#5};#6`#7`#8`{#9}]%                %      |       |
\enddiagram%                                       %      v       v
}}                                                 %     #4------>#5
\def\putptrianglep<#1>(#2,#3)[#4`#5`#6;#7`#8`#9]{{%
\settriparms[#1]%
\xpos=#2 \ypos=#3 \advance\ypos by \height
\puthmorphism(\xpos,\ypos)[#4`#5`{#7}]{\height}{\arrowtypea}a%
\putvmorphism(\xpos,\ypos)[`#6`{#8}]{\height}{\arrowtypeb}l%
\advance\xpos by\height
\putmorphism(\xpos,\ypos)(-1,-1)[``{#9}]{\height}{\arrowtypec}r%
}}

\def\putptriangle{\@ifnextchar <{\putptrianglep}{\putptrianglep
   <\arrowtypea`\arrowtypeb`\arrowtypec;\height>}}
\def\ptriangle{\@ifnextchar <{\ptrianglep}{\ptrianglep
   <\arrowtypea`\arrowtypeb`\arrowtypec;\height>}}
\def\ptrianglep<#1>[#2`#3`#4;#5`#6`#7]{{%%    %      #2----->#3
\settriparms[#1]%                             %      |      /
\diagram%                                     %      |     /
\putptrianglep<\arrowtypea`\arrowtypeb`%      %    #6|    /#7
\arrowtypec;\height>%                         %      |   /
(0,0)[#2`#3`#4;#5`#6`{#7}]%                   %      |  /
\enddiagram%%                                 %      v v
}}                                            %      #4

\def\putqtrianglep<#1>(#2,#3)[#4`#5`#6;#7`#8`#9]{{%
\settriparms[#1]%
\xpos=#2 \ypos=#3 \advance\ypos by\height
\puthmorphism(\xpos,\ypos)[#4`#5`{#7}]{\height}{\arrowtypea}a%
\putmorphism(\xpos,\ypos)(1,-1)[``{#8}]{\height}{\arrowtypeb}l%
\advance\xpos by\height
\putvmorphism(\xpos,\ypos)[`#6`{#9}]{\height}{\arrowtypec}r%
}}

\def\putqtriangle{\@ifnextchar <{\putqtrianglep}{\putqtrianglep
   <\arrowtypea`\arrowtypeb`\arrowtypec;\height>}}
\def\qtriangle{\@ifnextchar <{\qtrianglep}{\qtrianglep
   <\arrowtypea`\arrowtypeb`\arrowtypec;\height>}}
\def\qtrianglep<#1>[#2`#3`#4;#5`#6`#7]{{%%    %        #2----->#3
\settriparms[#1]%                             %         \      |
\width=\height                                %          \     |
\diagram%                                     %         #6\    |#7
\putqtrianglep<\arrowtypea`\arrowtypeb`%      %            \   |
\arrowtypec;\height>%                         %             \  |
(0,0)[#2`#3`#4;#5`#6`{#7}]%                   %              v v
\enddiagram%%                                 %               #4
}}

\def\putdtrianglep<#1>(#2,#3)[#4`#5`#6;#7`#8`#9]{{%
\settriparms[#1]%
\xpos=#2 \ypos=#3
\puthmorphism(\xpos,\ypos)[#5`#6`{#9}]{\height}{\arrowtypec}b%
\advance\xpos by \height \advance\ypos by\height
\putmorphism(\xpos,\ypos)(-1,-1)[``{#7}]{\height}{\arrowtypea}l%
\putvmorphism(\xpos,\ypos)[#4``{#8}]{\height}{\arrowtypeb}r%
}}

\def\putdtriangle{\@ifnextchar <{\putdtrianglep}{\putdtrianglep
   <\arrowtypea`\arrowtypeb`\arrowtypec;\height>}}
\def\dtriangle{\@ifnextchar <{\dtrianglep}{\dtrianglep
   <\arrowtypea`\arrowtypeb`\arrowtypec;\height>}}
\def\dtrianglep<#1>[#2`#3`#4;#5`#6`#7]{{%%    %                  / |
\settriparms[#1]%                             %                 /  |
\width=\height                                %              #5/   |#6
\diagram%                                     %               /    |
\putdtrianglep<\arrowtypea`\arrowtypeb`%      %              /     |
\arrowtypec;\height>%                         %             v      v
(0,0)[#2`#3`#4;#5`#6`{#7}]%                   %            #3----->#4
\enddiagram%%                                 %                #7
}}

\def\putbtrianglep<#1>(#2,#3)[#4`#5`#6;#7`#8`#9]{{%
\settriparms[#1]%
\xpos=#2 \ypos=#3
\puthmorphism(\xpos,\ypos)[#5`#6`{#9}]{\height}{\arrowtypec}b%
\advance\ypos by\height
\putmorphism(\xpos,\ypos)(1,-1)[``{#8}]{\height}{\arrowtypeb}r%
\putvmorphism(\xpos,\ypos)[#4``{#7}]{\height}{\arrowtypea}l%
}}

\def\putbtriangle{\@ifnextchar <{\putbtrianglep}{\putbtrianglep
   <\arrowtypea`\arrowtypeb`\arrowtypec;\height>}}
\def\btriangle{\@ifnextchar <{\btrianglep}{\btrianglep
   <\arrowtypea`\arrowtypeb`\arrowtypec;\height>}}
\def\btrianglep<#1>[#2`#3`#4;#5`#6`#7]{{%%   %              | \
\settriparms[#1]%                            %              |  \
\width=\height                               %            #5|   \#6
\diagram%                                    %              |    \
\putbtrianglep<\arrowtypea`\arrowtypeb`%     %              |     \
\arrowtypec;\height>%                        %              v      v
(0,0)[#2`#3`#4;#5`#6`{#7}]%                  %              #3----->#4
\enddiagram%%                                %                 #7
}}

\def\putAtrianglep<#1>(#2,#3)[#4`#5`#6;#7`#8`#9]{{%
\settriparms[#1]%
\xpos=#2 \ypos=#3 {\multiply \height by2
\puthmorphism(\xpos,\ypos)[#5`#6`{#9}]{\height}{\arrowtypec}b}%
\advance\xpos by\height \advance\ypos by\height
\putmorphism(\xpos,\ypos)(-1,-1)[#4``{#7}]{\height}{\arrowtypea}l%
\putmorphism(\xpos,\ypos)(1,-1)[``{#8}]{\height}{\arrowtypeb}r%
}}

\def\putAtriangle{\@ifnextchar <{\putAtrianglep}{\putAtrianglep
   <\arrowtypea`\arrowtypeb`\arrowtypec;\height>}}
\def\Atriangle{\@ifnextchar <{\Atrianglep}{\Atrianglep
   <\arrowtypea`\arrowtypeb`\arrowtypec;\height>}}
\def\Atrianglep<#1>[#2`#3`#4;#5`#6`#7]{{%%         %         /   \
\settriparms[#1]%                                  %        /     \
\width=\height                                     %     #5/       \#6
\diagram%                                          %      /         \
\putAtrianglep<\arrowtypea`\arrowtypeb`%           %     /           \
\arrowtypec;\height>%                              %    v             v
(0,0)[#2`#3`#4;#5`#6`{#7}]%                        %   #3------------>#4
\enddiagram%%                                      %          #7
}}

\def\putAtrianglepairp<#1>(#2)[#3;#4`#5`#6`#7`#8]{{%
\settripairparms[#1]%
\setpos(#2)%
\settokens`#3`%
\puthmorphism(\xpos,\ypos)[\tokenb`\tokenc`{#7}]{\height}{\arrowtyped}b%
\advance\xpos by\height
\puthmorphism(\xpos,\ypos)[\phantom{\tokenc}`\tokend`{#8}]%
{\height}{\arrowtypee}b%
\advance\ypos by\height
\putmorphism(\xpos,\ypos)(-1,-1)[\tokena``{#4}]{\height}{\arrowtypea}l%
\putvmorphism(\xpos,\ypos)[``{#5}]{\height}{\arrowtypeb}m%
\putmorphism(\xpos,\ypos)(1,-1)[``{#6}]{\height}{\arrowtypec}r%
}}

\def\putAtrianglepair{\@ifnextchar <{\putAtrianglepairp}{\putAtrianglepairp%
   <\arrowtypea`\arrowtypeb`\arrowtypec`\arrowtyped`\arrowtypee;\height>}}
\def\Atrianglepair{\@ifnextchar <{\Atrianglepairp}{\Atrianglepairp%
   <\arrowtypea`\arrowtypeb`\arrowtypec`\arrowtyped`\arrowtypee;\height>}}

\def\Atrianglepairp<#1>[#2;#3`#4`#5`#6`#7]{{%           %  #2a
\settripairparms[#1]%                         %           / | \
\settokens`#2`%                               %          /  |  \
\width=\height                                %       #3/  #4   \#5
\diagram%                                     %        /    |    \
\putAtrianglepairp                            %       /     |     \
<\arrowtypea`\arrowtypeb`\arrowtypec`%        %      v      v      v
\arrowtyped`\arrowtypee;\height>%             %     #2b---->#2c---->#2d
(0,0)[{#2};#3`#4`#5`#6`{#7}]%                 %         #6     #7
\enddiagram%%
}}

\def\putVtrianglep<#1>(#2,#3)[#4`#5`#6;#7`#8`#9]{{%
\settriparms[#1]%
\xpos=#2 \ypos=#3 \advance\ypos by\height {\multiply\height by2
\puthmorphism(\xpos,\ypos)[#4`#5`{#7}]{\height}{\arrowtypea}a}%
\putmorphism(\xpos,\ypos)(1,-1)[`#6`{#8}]{\height}{\arrowtypeb}l%
\advance\xpos by\height \advance\xpos by\height
\putmorphism(\xpos,\ypos)(-1,-1)[``{#9}]{\height}{\arrowtypec}r%
}}

\def\putVtriangle{\@ifnextchar <{\putVtrianglep}{\putVtrianglep
   <\arrowtypea`\arrowtypeb`\arrowtypec;\height>}}
\def\Vtriangle{\@ifnextchar <{\Vtrianglep}{\Vtrianglep
   <\arrowtypea`\arrowtypeb`\arrowtypec;\height>}}
\def\Vtrianglep<#1>[#2`#3`#4;#5`#6`#7]{{%%     %        #2------------->#3
\settriparms[#1]%                              %         \             /
\width=\height                                 %          \           /
\diagram%                                      %         #6\         /#7
\putVtrianglep<\arrowtypea`\arrowtypeb`%       %            \       /
\arrowtypec;\height>%                          %             \     /
(0,0)[#2`#3`#4;#5`#6`{#7}]%                    %              v   v
\enddiagram%%                                  %               #4
}}

\def\putVtrianglepairp<#1>(#2)[#3;#4`#5`#6`#7`#8]{{
\settripairparms[#1]%
\setpos(#2)%
\settokens`#3`%
\advance\ypos by\height
\putmorphism(\xpos,\ypos)(1,-1)[`\tokend`{#6}]{\height}{\arrowtypec}l%
\puthmorphism(\xpos,\ypos)[\tokena`\tokenb`{#4}]{\height}{\arrowtypea}a%
\advance\xpos by\height
\puthmorphism(\xpos,\ypos)[\phantom{\tokenb}`\tokenc`{#5}]%
{\height}{\arrowtypeb}a%
\putvmorphism(\xpos,\ypos)[``{#7}]{\height}{\arrowtyped}m%
\advance\xpos by\height
\putmorphism(\xpos,\ypos)(-1,-1)[``{#8}]{\height}{\arrowtypee}r%
}}

\def\putVtrianglepair{\@ifnextchar <{\putVtrianglepairp}{\putVtrianglepairp%
    <\arrowtypea`\arrowtypeb`\arrowtypec`\arrowtyped`\arrowtypee;\height>}}
\def\Vtrianglepair{\@ifnextchar <{\Vtrianglepairp}{\Vtrianglepairp%
    <\arrowtypea`\arrowtypeb`\arrowtypec`\arrowtyped`\arrowtypee;\height>}}
\def\Vtrianglepairp<#1>[#2;#3`#4`#5`#6`#7]{{%  %  #2a---->#2b---->#2c
\settripairparms[#1]%                          %   \      |      /
\settokens`#2`%                                %    \     |     /
\diagram%                                      %   #5\   #6    /#7
\putVtrianglepairp                             %      \   |   /
<\arrowtypea`\arrowtypeb`\arrowtypec`%         %       \  |  /
\arrowtyped`\arrowtypee;\height>%              %        v v v
(0,0)[{#2};#3`#4`#5`#6`{#7}]%                  %         #2d
\enddiagram%%
}}

\def\putCtrianglep<#1>(#2,#3)[#4`#5`#6;#7`#8`#9]{{%
\settriparms[#1]%
\xpos=#2 \ypos=#3 \advance\ypos by\height
\putmorphism(\xpos,\ypos)(1,-1)[``{#9}]{\height}{\arrowtypec}l%
\advance\xpos by\height \advance\ypos by\height
\putmorphism(\xpos,\ypos)(-1,-1)[#4`#5`{#7}]{\height}{\arrowtypea}l%
{\multiply\height by 2
\putvmorphism(\xpos,\ypos)[`#6`{#8}]{\height}{\arrowtypeb}r}%
}}

\def\putCtriangle{\@ifnextchar <{\putCtrianglep}{\putCtrianglep
    <\arrowtypea`\arrowtypeb`\arrowtypec;\height>}}
\def\Ctriangle{\@ifnextchar <{\Ctrianglep}{\Ctrianglep
    <\arrowtypea`\arrowtypeb`\arrowtypec;\height>}}
\def\Ctrianglep<#1>[#2`#3`#4;#5`#6`#7]{{%%   %                / |
\settriparms[#1]%                            %             #5/  |
\width=\height                               %              /   |
\diagram%                                    %             v    |
\putCtrianglep<\arrowtypea`\arrowtypeb`%     %           #3     |#6
\arrowtypec;\height>%                        %             \    |
(0,0)[#2`#3`#4;#5`#6`{#7}]%                  %            #7\   |
\enddiagram%%                                %               \  |
}}                                           %                v v
\def\putDtrianglep<#1>(#2,#3)[#4`#5`#6;#7`#8`#9]{{%
\settriparms[#1]%
\xpos=#2 \ypos=#3 \advance\xpos by\height \advance\ypos by\height
\putmorphism(\xpos,\ypos)(-1,-1)[``{#9}]{\height}{\arrowtypec}r%
\advance\xpos by-\height \advance\ypos by\height
\putmorphism(\xpos,\ypos)(1,-1)[`#5`{#8}]{\height}{\arrowtypeb}r%
{\multiply\height by 2
\putvmorphism(\xpos,\ypos)[#4`#6`{#7}]{\height}{\arrowtypea}l}%
}}

\def\putDtriangle{\@ifnextchar <{\putDtrianglep}{\putDtrianglep
    <\arrowtypea`\arrowtypeb`\arrowtypec;\height>}}
\def\Dtriangle{\@ifnextchar <{\Dtrianglep}{\Dtrianglep
   <\arrowtypea`\arrowtypeb`\arrowtypec;\height>}}
\def\Dtrianglep<#1>[#2`#3`#4;#5`#6`#7]{{%%  %          | \
\settriparms[#1]%                           %          |  \#6
\width=\height                              %          |   \
\diagram%                                   %          |    v
\putDtrianglep<\arrowtypea`\arrowtypeb`%    %        #5|    #3
\arrowtypec;\height>%                       %          |    /
(0,0)[#2`#3`#4;#5`#6`{#7}]%                 %          |   /#7
\enddiagram%%                               %          |  /
}}                                          %          v v
\def\setrecparms[#1`#2]{\width=#1 \height=#2}%
\def\recursep<#1`#2>[#3;#4`#5`#6`#7`#8]{{\m@th
\width=#1 \height=#2 \settokens`#3`
\settowidth{\tempdimen}{$\tokena$} \ifdim\tempdimen=0pt
  \savebox{\tempboxa}{\hbox{$\tokenb$}}%
  \savebox{\tempboxb}{\hbox{$\tokend$}}%
  \savebox{\tempboxc}{\hbox{$#6$}}%
\else
  \savebox{\tempboxa}{\hbox{$\hbox{$\tokena$}\times\hbox{$\tokenb$}$}}%
  \savebox{\tempboxb}{\hbox{$\hbox{$\tokena$}\times\hbox{$\tokend$}$}}%
  \savebox{\tempboxc}{\hbox{$\hbox{$\tokena$}\times\hbox{$#6$}$}}%
\fi \ypos=\height \divide\ypos by 2 \xpos=\ypos \advance\xpos by
\width \bfig
\putCtrianglep<-1`1`1;\ypos>(0,0)[`\tokenc`;#5`#6`{#7}]%
\puthmorphism(\ypos,0)[\tokend`\usebox{\tempboxb}`{#8}]{\width}{-1}b%
\puthmorphism(\ypos,\height)[\tokenb`\usebox{\tempboxa}`{#4}]{\width}{-1}a%
\advance\ypos by \width
\putvmorphism(\ypos,\height)[``\usebox{\tempboxc}]{\height}1r%
\efig }}

\def\recurse{\@ifnextchar <{\recursep}{\recursep<\width`\height>}}

\def\puttwohmorphisms(#1,#2)[#3`#4;#5`#6]#7#8#9{{%
\puthmorphism(#1,#2)[#3`#4`]{#7}0a \ypos=#2 \advance\ypos by 20
\puthmorphism(#1,\ypos)[\phantom{#3}`\phantom{#4}`#5]{#7}{#8}a
\advance\ypos by -40
\puthmorphism(#1,\ypos)[\phantom{#3}`\phantom{#4}`#6]{#7}{#9}b }}

\def\puttwovmorphisms(#1,#2)[#3`#4;#5`#6]#7#8#9{{%
\putvmorphism(#1,#2)[#3`#4`]{#7}0a \xpos=#1 \advance\xpos by -20
\putvmorphism(\xpos,#2)[\phantom{#3}`\phantom{#4}`#5]{#7}{#8}l
\advance\xpos by 40
\putvmorphism(\xpos,#2)[\phantom{#3}`\phantom{#4}`#6]{#7}{#9}r }}

\def\puthcoequalizer(#1)[#2`#3`#4;#5`#6`#7]#8#9{{%
\setpos(#1)%
\puttwohmorphisms(\xpos,\ypos)[#2`#3;#5`#6]{#8}11%
\advance\xpos by #8
\puthmorphism(\xpos,\ypos)[\phantom{#3}`#4`#7]{#8}1{#9} }}

\def\putvcoequalizer(#1)[#2`#3`#4;#5`#6`#7]#8#9{{%
\setpos(#1)%
\puttwovmorphisms(\xpos,\ypos)[#2`#3;#5`#6]{#8}11%
\advance\ypos by -#8
\putvmorphism(\xpos,\ypos)[\phantom{#3}`#4`#7]{#8}1{#9} }}

\def\putthreehmorphisms(#1)[#2`#3;#4`#5`#6]#7(#8)#9{{%
\setpos(#1) \settypes(#8)
\if a#9 %
     \vertsize{\tempcounta}{#5}%
     \vertsize{\tempcountb}{#6}%
     \ifnum \tempcounta<\tempcountb \tempcounta=\tempcountb \fi
\else
     \vertsize{\tempcounta}{#4}%
     \vertsize{\tempcountb}{#5}%
     \ifnum \tempcounta<\tempcountb \tempcounta=\tempcountb \fi
\fi \advance \tempcounta by 60
\puthmorphism(\xpos,\ypos)[#2`#3`#5]{#7}{\arrowtypeb}{#9}
\advance\ypos by \tempcounta
\puthmorphism(\xpos,\ypos)[\phantom{#2}`\phantom{#3}`#4]{#7}{\arrowtypea}{#9}
\advance\ypos by -\tempcounta \advance\ypos by -\tempcounta
\puthmorphism(\xpos,\ypos)[\phantom{#2}`\phantom{#3}`#6]{#7}{\arrowtypec}{#9}
}}

\def\setarrowtoks[#1`#2`#3`#4`#5`#6]{%
\def\toka{#1}
\def\tokb{#2}
\def\tokc{#3}
\def\tokd{#4}
\def\toke{#5}
\def\tokf{#6}
}
\def\hex{\@ifnextchar <{\hexp}{\hexp<1000`400>}}
\def\hexp<#1`#2>[#3`#4`#5`#6`#7`#8;#9]{%
\setarrowtoks[#9] \yext=#2 \advance \yext by #2 \xext=#1
\advance\xext by \yext \bfig
\putCtriangle<-1`0`1;#2>(0,0)[`#5`;\tokb``\tokd] \xext=#1 \yext=#2
\advance \yext by #2
\putsquare<1`0`0`1;\xext`\yext>(#2,0)[#3`#4`#7`#8;\toka```\tokf]
\advance \xext by #2
\putDtriangle<0`1`-1;#2>(\xext,0)[`#6`;`\tokc`\toke] \efig }
\makeatother